\documentclass[aps,twocolumn,superscriptaddress]{revtex4}

\usepackage{graphicx}
\usepackage{dcolumn}
\usepackage{bm,gensymb,multirow}
\usepackage{lipsum}
\usepackage{natbib,amsmath}

\usepackage[T1]{fontenc}

\usepackage{xcolor}

\newcommand{\pwr}
{\affiliation{Department of Operations Research and Business Intelligence, Politechnika Wroclawska, Wybrzeze Stanislawa Wyspianskiego 27, 50370 Wroclaw, Lower Silesia, Poland}}

\newcommand{\cu}{\affiliation{Department of Physics, University of Calcutta, 92 Acharya Prafulla Chandra Road, Kolkata 700009, West Bengal, India}}

\begin{document}


\title{Analysing contrarian behaviour using nonlinear biased $q$-voter model}


\author{Amit Pradhan}
\cu

\author{Pratik Mullick}
\pwr

\author{Parongama Sen}
\cu


\date{\today}

\begin{abstract}
We investigate the role of contrarians in a recently proposed weighted-influence variant of the $q$-voter model. In this framework, non-unanimous influence groups affect the focal agent through weighted contributions governed by a bias parameter $p$. We extend this setting by introducing a fraction $\alpha$ ($\alpha> 0$) of contrarians, defined as agents who systematically oppose the prevailing influence irrespective of whether the group is unanimous or divided. Analytical mean-field calculations and Monte Carlo simulations reveal that the final states of the system are governed by simple phase boundaries: regions of positive and negative majority separated by the lines $p=1/2$ and $\alpha=1/2$, with equally-mixed states confined to these boundaries. While low contrarian densities are insufficient to overturn the bias, higher values of $\alpha$ systematically drive the system closer to a balanced coexistence of opinions, though exact parity is prevented by the presence of bias $p$. We further analyze the temporal relaxation of opinions and extract the characteristic timescales of convergence. Our findings highlight how contrarians, acting as structured non-conformists, can suppress consensus and maintain opinion diversity, while internal biases ultimately hinder a perfectly even split.
\end{abstract}


\maketitle


\section{Introduction}

Understanding how collective decisions emerge from individual interactions is a central question in both physics-inspired modeling of social systems and the social sciences. In the framework of sociophysics, minimal models of opinion dynamics \cite{galam2008sociophysics,castellano2009statistical,sen2014sociophysics,sznajd2021review,jusup2022social} have proven to be useful tools for understanding the mechanisms that drive consensus formation, polarization, and coexistence in populations. These models simplify the complexities of human behavior while retaining the key elements of collective dynamics, allowing for both analytical treatment and clear interpretation of results. Such approaches have found applications in diverse domains, including political elections \cite{gsanger2024opinion}, financial markets \cite{zha2020opinion}, innovation adoption \cite{martins2009opinion}, and the spread of cultural norms \cite{castellano2009statistical}.

Binary decision-making models, where each agent chooses between two discrete alternatives, constitute one of the most studied classes of opinion dynamics models. They provide a simple yet effective representation of situations in which individuals must align with or oppose a given choice, such as voting for one of the two candidates, supporting or rejecting a policy, or deciding to adopt or forgo a technology etc. The original voter model \cite{holley1975ergodic,liggett1985interacting} represents perhaps the simplest case of binary opinion dynamics. In this framework, an agent adopts the opinion of a randomly chosen neighbor.  Owing to its simplicity and capacity to study core mechanisms of opinion change in social systems, the voter model has become one of the most widely studied frameworks in the field \cite{redner2019reality}, giving rise to several extensions. One such extension is the $q$-voter model \cite{castellano2009nonlinear}, also referred to as the nonlinear voter model, which is now considered a standard framework for studying collective behaviour of opinion formation. This model generalizes the interaction rule by allowing a group of $q$ neighbors (the $q$-panel) to jointly influence the focal agent. If the panel is unanimous, the focal agent adopts its opinion, a behaviour known as conformity; otherwise, the agent changes its opinion with a certain probability, representing stochasticity or independent behavior.

Over the years, the $q$-voter model has been further extended in numerous ways to study a wider variety of social influence mechanisms. Several variants combine conformity with responses such as independence \cite{nyczka2012phase,byrka2016difficulty}, anti-conformity \cite{nyczka2013anticonformity,javarone2015conformism}, an agent’s resistance to persuasion \cite{anugraha2025nonlinear}, while others introduce ``zealots" or inflexible agents that stick to their opinion  \cite{mobilia2015nonlinear}. The model has also been implemented on diverse network structures, including lattices \cite{timpanaro2014exit}, multiplex networks \cite{gradowski2020pair}, Barabasi-Albert networks \cite{fardela2025opinion}, duplex cliques \cite{chmiel2015phase}, random graphs \cite{gradowski2020pair} and scale-free networks \cite{vieira2020pair,anugraha2025nonlinear}. Another variant, namely the threshold $q$-voter model \cite{vieira2018threshold,nyczka2018conformity,vieira2020pair}, relax the unanimity condition. In this case unanimity among minimum $q_0$ number of agents in the $q$-panel $(0<q_0\leq q)$ is considered enough to sway the focal agent. In the case of a non-unanimous $q$-panel, more recent adaptions study the effect of mass media influence \cite{muslim2024mass,fardela2025opinion}, weighted influence mechanisms \cite{mullick2025social}, and susceptibility \cite{civitarese2021external} to unreliable influence mechanism. Overall, these extensions illustrate the model’s adaptability to a range of social dynamics.

In the context of opinion dynamics, contrarians are agents who deliberately adopt the opposite stance to the majority opinion in their local neighborhood. Representing skeptical individuals, ideological opponents, or loyal minority consumers, they form a systematic type of nonconformity, distinct from random independence, and are known to hinder full consensus and preserve opinion diversity. Being a contrarian was originally regarded as an investment strategy in financial markets \cite{lo1990contrarian,corcos2002imitation,park2011herding}. The incorporation of contrarians has also been shown \cite{zhang2016suppressing} to suppress explosive synchronization, a phenomenon linked to cascading power-grid failures and epileptic seizures.

Galam in 2004 \cite{galam2004contrarian} first explored the role of contrarians in the opinion dynamics of human societies using majority rule model. The  contrarian fraction here could be tuned to show the existence of a phase transition: below a critical threshold of this fraction, a polarized phase with a clear majority exists, while above it the system reaches a ``hung'' state with equal support for both opinions. This pioneering model was formulated in an annealed setting.
A corresponding quenched implementation, where a fixed fraction of agents are permanently contrarian, was later investigated in \cite{stauffer2004simulation}, yielding a similar transition. Further extensions examined proportionate contrarians, whose tendency to oppose the majority increases with its size, and one-sided contrarians, who systematically oppose only one of the two opinions \cite{borghesi2006chaotic}. Subsequently a dynamical phase transition was demonstrated in systems containing both contrarians and opportunists (agents who always adopt the local majority opinion) \cite{kurten2008dynamical}.

Since Galam’s original work on the majority-rule model, the impact of contrarians has been explored in a wide range of other dynamical settings, including the Sznajd model \cite{schneider2004influence,de2005spontaneous,de2006van,de2021contrarian}, minority games \cite{zhong2005effects}, Kuramoto oscillators \cite{hong2011conformists,manoranjani2024asymmetric}, continuous-opinion models \cite{martins2010importance}, antiferromagnetic Ising models \cite{bagnoli2013topological}, and kinetic exchange processes with three states \cite{crokidakis2014impact,gambaro2017influence}. Within the majority-rule framework itself, contrarians have also been studied in systems with mobile agents \cite{guo2015opinion}, in the presence of zealots \cite{jacobs2019two}, under external propaganda fields \cite{gimenez2023contrarian}, and incorporating local biased tie-breaking \cite{galam2025democratic}. These models have been implemented on various topologies including lattices \cite{schneider2004influence,de2005spontaneous,de2006van,martins2010importance,guo2015opinion}, Watts–Strogatz networks \cite{bagnoli2013topological,de2021contrarian}, Erdos–Renyi and Barabási–Albert graphs \cite{de2021contrarian}, and complete graphs \cite{crokidakis2014impact,gambaro2017influence,jacobs2019two,gimenez2023contrarian}. Across these studies, contrarians have consistently been shown to hinder the formation of a clear majority, although a ``tit-for-tat" strategy has been proposed to counteract this effect and restore consensus \cite{zhang2013efficient}.

Since the 2010s, contrarian behavior has also been incorporated into various voter and voter-like models \cite{masuda2013voter,tanabe2013complex,yi2013phase,banisch2014microscopic,khalil2019noisy}. In contrast, studies of contrarians within the $q$-voter framework remain relatively scarce \cite{nyczka2012phase,nyczka2013anticonformity,anugraha2025nonlinear}. Nyczka et al. \cite{nyczka2012phase} examined a model in which, with probability $\alpha$, a focal agent behaves as an anti-conformist contrarian whenever the $q$-panel is unanimous. As in previous models, they observed a continuous phase transition: for small values of $\alpha$, the system reaches a polarized state, whereas larger $\alpha$ suppresses the emergence of majority. The critical value of $\alpha$ increases with $q$ but saturates for large $q$. In a follow-up work \cite{nyczka2013anticonformity}, they introduced a variant in which contrarian behavior occurs only if at least $r$ members of the $q$-panel agree, again finding a continuous phase transition with a critical $\alpha$ that grows with $q$. More recently, Anugraha et al. \cite{anugraha2025nonlinear} studied anti-conformist contrarians on a Barabási–Albert network and reported similar continuous transitions.

In this paper, we explore a recent variant of the $q$-voter model \cite{mullick2025social} by introducing a fraction $\alpha$ of contrarians and studying its effect on the system’s final state. Earlier, in this model without contrarians, it was considered the agents change their opinions according to
the actual configuration of the $q$-panel when it is not unanimous, with a bias towards positive opinion governed by a parameter $p$. It was found that the final states are either consensus states (for the unbiased case $p \neq 1/2$), or states with  mixed opinions only for $p=1/2$. 
Building on this setup, we define contrarians to act against both unanimous influence and the weighted influence of non-unanimous groups, thereby extending the mechanism of opposition. 
Unlike earlier works on anti-conformist contrarians, where agents oppose social influence only when the $q$-panel is unanimous \cite{nyczka2012phase,nyczka2013anticonformity,anugraha2025nonlinear}, the contrarians in this model oppose the influence of their neighbors irrespective of whether the $q$-panel is unanimous or not. 

Owing to this difference, our results differ markedly from previous studies: emergence of a mixed phase with equal support for both the opinions takes place only when either $\alpha=1/2$ (and any value of $p$) or $p=1/2$ (and any value of $\alpha$). In all other cases, the system is polarized with a clear majority. Even for higher values of fraction $\alpha$ of contrarians we do not get a ``hung'' election situation, contrary to the results in the existing literature.  We also obtain a rich phase diagram for the steady state value of fraction of positive opinion as a function of $p$ and $\alpha$. These results show small dependence on $q$ for its small values, but becomes $q$ independent as $q$ increases. In addition, we compute the characteristic time-scale $\tau$ associated with the temporal evolution of the fraction of positive agents, analysing its dependence on $q$, $\alpha$ and $p$.

The rest of the paper is structured as follows: in section \ref{sec:model} we define our model, describe its underlying dynamical rule and briefly mention the methods that have been used to the study the model. Then in section \ref{sec:res_dis} we summarize our key findings and their interpretations. We end by some concluding remarks in section \ref{sec: conclu}.

\section{Model description and methods used}\label{sec:model}

We consider a population of $N$ agents with binary opinions, either positive or negative, interacting on a fully connected network. The system evolves over discrete time steps, and at each step, a randomly selected agent $A$, called the focal agent, interacts with a group of $q\geq 2$ other agents, called a $q$-panel. This $q$-panel acts as the influence group for the focal agent. The opinion dynamics for the focal agent in our model is two-fold and could be described as following:


\begin{itemize}
    \item[(a)] If the $q$ panel is unanimous, then the focal agent takes that opinion. This is the case for non-contrarians, the contrarian agent will do exactly the opposite.
    
    \item[(b)] If on the other hand, the $q$ panel is non-unanimous, then the focal agent considers the influential power of each subgroup in the panel. We consider, in the $q$ panel, each positive agent has an influential power $p$ and each negative agent has an influential power $1-p$. Let there be $n$ agents with positive opinion in that panel. These $n$ agents with influential power $p$ each, would try to convince $A$ with a total influential power of $np$. Quite similarly, $q-n$ agents with negative opinion each with influential power $1-p$, would have a total influential power of $(q-n)(1-p)$ on $A$. Here we consider that the effective probabilities with which $A$ should choose either of the opinions are weighted averages of the influential powers of positive and negative agents. So if $p_{q+}$ and $p_{q-}$ denote the probabilities that $A$ should take the positive opinion and the negative opinion respectively, then \begin{eqnarray}
    p_{q+} = \frac{np}{np + (q-n)(1-p)}\label{p_q+}\\[0.2cm]
    p_{q-} = \frac{(q-n)(1-p)}{np+(q-n)(1-p)}\label{p_q-}
    \end{eqnarray} Clearly the denominator in the above equations $np+(q-n)(1-p)$ is the normalisation factor. 
    Once again, the contrarian will take the opposite opinion as 
    contrasted to the ordinary agent.  \cite{mullick2025social}. 
\end{itemize}


In this study, we consider a finite fraction $\alpha$ of agents who act as contrarians. $\alpha = 0$, the case with no contrarians had been studied in \cite{mullick2025social} and our discussions here are for $\alpha >0$. 
The selection of the contrarians  can be made in two ways: (i) quenched approach – at the start of each configuration,  $\alpha$ fraction of agents are designated as contrarians and retain this behavior throughout the simulation, or (ii) annealed approach – at each microscopic opinion update, the focal agent behaves as a contrarian with probability $\alpha$. Our numerical simulations show that both approaches yield identical results.

Let $f_+(t)$ denote the fraction of agents with positive opinion within the entire system at time $t$. Then $f_+(t)$ also represents the probability that any randomly selected agent has a positive opinion at time $t$. Naturally $f_-(t) = 1-f_+(t)$ is considered to be the probability of a negative opinion. The initial fraction of agents with positive opinion is denoted by $f_0 = f_+(t=0)$ throughout this paper. We use mean field theory (valid for $N \to \infty$) to obtain the dynamical equations governing $f_+(t)$ for different values of $q$ and solve the equations analytically for $q=2$ and numerically for other values of $q$. We also make a fixed point analysis for small $q$ values as well as for $q\to \infty$.

We complemented the mean field analysis with Monte Carlo (MC) simulations to study the temporal evolution of $f_+(t)$ and compare it with theoretical predictions. The simulations start with a system of $N$ agents, of which a fraction $f_0$ $(=f_+(0))$ initially hold the positive opinion. Among these, a fraction $\alpha$ are designated as contrarians, meaning they adopt the opposite opinion from the one prescribed by the update rule. We employ a random asynchronous update scheme: at each step, a focal agent is selected uniformly at random and updated immediately according to the defined dynamical rules. One MC time step consists of $N$ such updates. The $q$-panel is selected using the Fisher–Yates algorithm \cite{fisher1953statistical}, ensuring random sampling without repetition and excluding the focal agent from the panel. Simulations proceed until a predetermined maximum number of MC steps is reached, and the resulting $f_+(t)$ is averaged over multiple independent realizations with different initial configurations.

In the following section, we present the mean-field analysis and compare its predictions with numerical results obtained from Monte Carlo simulations.

\section{Results and Discussion}\label{sec:res_dis}

\subsection{Mean field approach}

We consider a fully connected network of $N$ agents, which allows the application of mean field theory. In this framework, the master equation for $f_{+}(t)$, the probability that an agent holds a positive opinion, can be expressed in terms of the transition rates between the two opinion states. With $\alpha$ denoting the probability that an agent acts as a contrarian, the transition rates $\omega$ from the positive ($+$) state to the negative ($-$) state, and vice versa, are given by
\begin{equation}
\label{negative to positive rate}
\begin{split}
\omega_{-\to+} = (1-\alpha)\left [f_+^{q} + \sum_{n=1}^{q-1} p_{q+} \binom {q}{n}f_{+}^{n}(1-f_+)^{q-n}\right]\\
+ \alpha \left [(1-f_+)^{q} + \sum_{n=1}^{q-1}p_{q-} \binom{q}{n} f_+^n(1-f_+)^{q-n}\right]
\end{split}
\end{equation}
\begin{equation}
\label{positive to negative rate}
\begin{split}
\omega_{+\to-} = (1-\alpha)\left [(1-f_+)^{q} + \sum_{n=1}^{q-1} p_{q-} \binom {q}{n}f_{+}^{n}(1-f_+)^{q-n}\right]\\ 
+ \alpha \left [f_+^{q} + \sum_{n=1}^{q-1}p_{q+} \binom{q}{n} f_+^n(1-f_+)^{q-n}\right].
  \end{split}
\end{equation}
Eq. (\ref{negative to positive rate}) expresses the transition rate $\omega_{-\to+}$ from a negative state to a positive state, where inside the brackets on the right hand side, the first term corresponds to conformity, i.e., all the $q$ agents have the same opinion (either positive or negative with a probability $(1-\alpha)$ or $\alpha$ respectively) and the second term corresponds to all other cases corresponding to probability $1-\alpha$ and $\alpha$. Eq. (\ref{positive to negative rate}) is very similar to Eq. (\ref{negative to positive rate}) except that it expresses the transition rate $\omega_{+\to-}$ from a negative state to a positive state.

The master equation for $f_+(t)$ could then be written as
\begin{equation}
\label{master equation}
 \frac{df_{+}(t)}{dt} = -\omega_{+\to-}f_{+}(t)+\omega_{-\to +}f_{-}(t)
\end{equation}
such that one gets on simplification
\begin{equation}
\label{simplified master equation}
 \frac{df_+(t)}{dt} = (1-2\alpha)\left[\sum_{n=0}^{q}p_{q+} \binom{q}{n} f_+^{n}(1-f_+)^{q-n}\right] + \alpha - f_+
\end{equation}

{\textit {Specific cases}}:\\
We note that in Eq. (\ref{simplified master equation}), putting $\alpha = 1/2$ produces
${df_+}/{dt} = 1/2-f_+$ irrespective of the value of $q$. This suggests that for $\alpha = 1/2$, for any value of $q$, there is only one fixed point $f_+^{*} = 1/2$ independent of the parameter $p$.  We consider an infinitesimal deviation $\delta$ from the fixed point  by putting $f_+ = f_+^{*} + \delta$ to 
 get
\begin{equation}
\label{equation for delta}
 \frac{d\delta}{dt} = -\delta,
\end{equation}
which has a solution $\delta\sim e^{-t}$. This clearly suggests that the flow converges towards the fixed point $f_+^{*} = 1/2$, and thus we conclude that the fixed point $f_+^{*} = 1/2$ is asymptotically stable which is pretty obvious as there is only one fixed point.

Also if we put $p=1/2$ in Eq. (\ref{simplified master equation}), then with $p_{q+}=n/q$ we get
\begin{equation}
\label{rate equation involving <n>}
\frac{df_+}{dt} = \frac{(1-2\alpha)}{q}\left[\sum_{n=0}^{q} n \binom{q}{n} f_+(1-f_+)^{q-n}\right]+\alpha-f_+.
\end{equation}
The term within the parentheses on the right hand side is simply the average value of the random variable $n$ which is $qf_+$. Therefore Eq.\ref{rate equation involving <n>} simplifies to
\begin{equation}
\label{simplified rate equation involving <n>}
\frac{df_+}{dt} = \alpha(1-2f_+).
\end{equation}
This is exactly the equation followed by the mean field voter model
with contrarians. This is not surprising, as for $p=1/2$,  in absence of contrarians,  the model reduces to the mean field voter model.
Again, $f_+=1/2$ is a fixed point irrespective of the value of $\alpha$ except for $\alpha=0$.

A linear stability analysis shows that the deviation $\delta $ from the fixed point follows the behavior
 \begin{equation}
  \frac{d\delta}{dt} = -2 \alpha \delta.   
 \end{equation}
This has a solution $\delta\sim e^{-2 \alpha t}$,  showing that the flow towards the stable fixed point $f_+^{*} = 1/2$ is dependent on the value of $\alpha$. 

If we transform $p$ and $f_+$ in such a way that $p\to 1-p$ and $f_+ \to 1-f_+$, then it turns out by simple algebra that the dynamical equation for $f_+ $ [Eq. (\ref{simplified master equation})] will remain invariant under the above set of transformation in, absence of $\alpha$ as well as in presence of $\alpha$, for all values of $q$. So the symmetry would persist even if we include $\alpha$.

The first term on the right-hand side of Eq. (\ref{simplified master equation}) contains a sum that becomes cumbersome to evaluate for large values of $q$. For small $q$, however, the terms can be computed explicitly to examine the dynamical evolution of $f_+$. In the following, we present the analysis for $q = 2$ and $q = 3$. A simplified mean field theory, best applicable for $q \to \infty$ has also been presented in section \ref{simpmf}. Of course, numerical simulations have been done for larger $q$ and as we show later, the qualitative results are quite similar as one spans $q$ values from 2  to $\infty$.

Remarkably, for any $q$, there is only one fixed point for $f_+$ in general, for any $\alpha$ and $p$, which is obviously a stable  fixed point lying between 0 and 1. Thus there is no scope for the existence of a chaotic or disordered regime. 

\subsubsection{q = 2 case}

We discuss the $q=2$ case in greater detail as it can be handled analytically to get a number of interesting results. 
For $q=2$,  Eq. (\ref{simplified master equation}), reduces to
\begin{equation}
\label{Simplified master equation for q=2}  
\frac{df_+}{dt} = Af_{+}^{2} + Bf_+ + \alpha,
\end{equation} 
where $A=4\alpha p-2\alpha-2p+1,B = 2p-1-4\alpha p$ and  we define $\Delta = B^2-4A\alpha$. Upon solving the differential equation given by Eq. (\ref{Simplified master equation for q=2}) analytically subjected to the initial condition $f_+(0)=f_0=0$, we obtain a closed form solution (see Eq. (\ref{exact expression of f_+ for q=2}) of Appendix A)  as 
\begin{equation}
\label{Exact expression of f_+ for q=2}
    f_+(t) = \frac{\sqrt{\Delta}}{2A}\left[\frac{1+\left(\frac{B-\sqrt{\Delta}}{B+\sqrt{\Delta}}\right)e^{\sqrt{\Delta}t}}{1-\left(\frac{B-\sqrt{\Delta}}{B+\sqrt{\Delta}}\right)e^{\sqrt{\Delta}t}}\right]-\frac{B}{2A}.
\end{equation}.

Hence, one can obtain $f_+(t)$ in the limit $t\to\infty$ as 
$f_+(\infty)=f_+^* = -(\sqrt{\Delta}+B)/2A$ for all $\alpha$ and $p$ except  for $\alpha = 0.5$ and/or $p=0.5$ for which $A =0$ and the above solution is invalid.
One can directly get the solution for $\alpha = 0.5$ as $f_+(t) = \frac{1}{2} (1-e^{-t})$  for all $p$ and for $p=0.5$, as long as $\alpha$ is non zero, $f_+(t) = \frac{1}{2}(1-e^{-2\alpha t})$. For both cases   $f_+(\infty) = 1/2$
as already discussed for any $q$.  
It may be recalled that for $p=0.5,\alpha=0$, any point is a fixed point \cite{mullick2025social}. 

\begin{figure*}
    \centering
    \includegraphics[width=\textwidth]{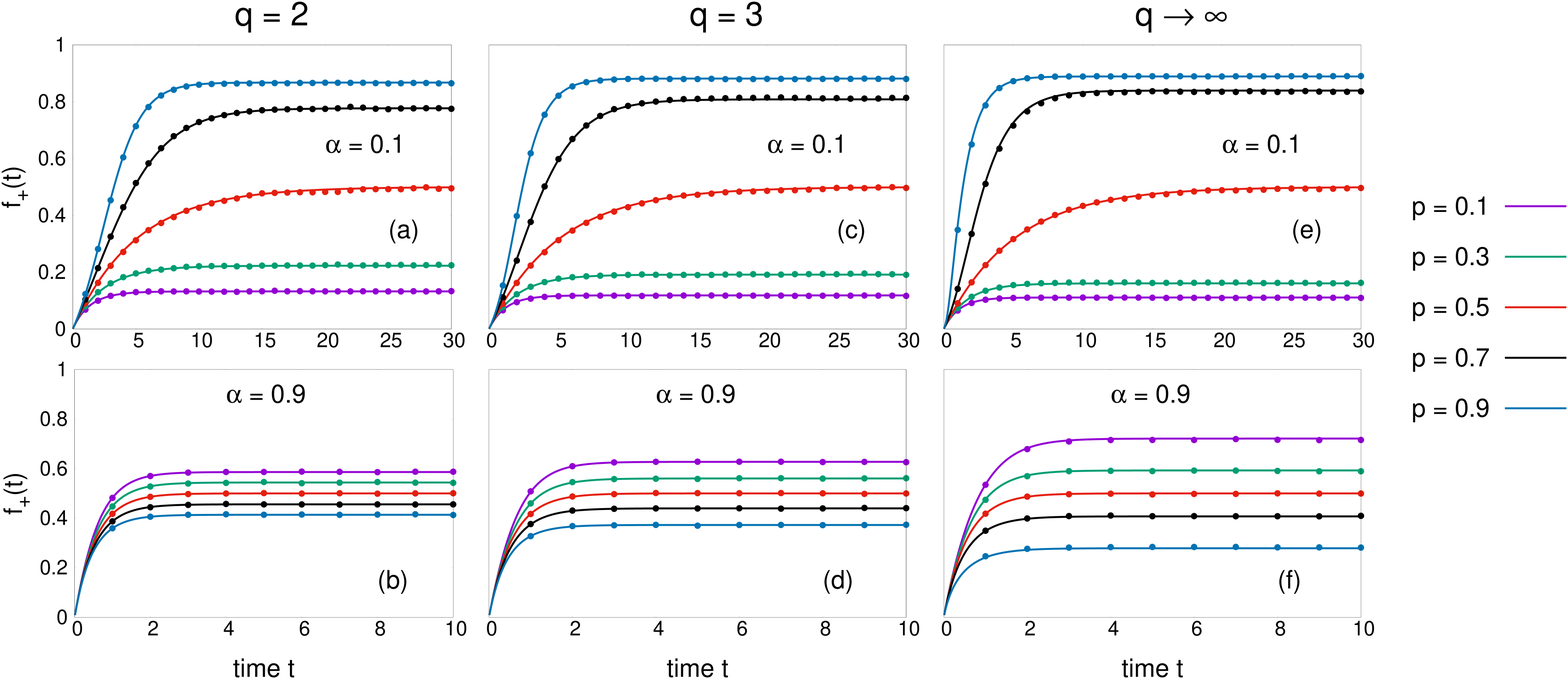}
    \caption{Variations of fraction $f_+(t)$ of agents with positive opinion as a function of time $t$ for several values of $q$ and $\alpha$, such as (a) $q=2$, $\alpha=0.1$ (b) $q=2$, $\alpha=0.9$, (c) $q=3$, $\alpha=0.1$, (d) $q=3$, $\alpha=0.9$, (e) $q\rightarrow\infty$, $\alpha=0.1$ and (f) $q\rightarrow\infty$, $\alpha=0.9$. Plots are shown for $p=0.1,0.3,0.5,0.7,0.9$ and for an initial condition $f_0 = 0$. Here, analytical and simulated results are shown by solid lines and solid circles respectively. Analytical results for $q\rightarrow\infty$ are compared with simulated results for $q=50$ and the agreement seems excellent. Simulations were performed using a system size of $N=2^{10}$ averaging over $10^2$ configurations.}
    \label{trajectory for q=2,3,infty}
\end{figure*}


 In Fig. \ref{trajectory for q=2,3,infty}, $f_+$ as a function of time $t$, obtained by both mean field calculations and numerical simulations, is plotted for two values of fractions of contrarians $\alpha$. The agreement between the theoretical and numerical results is excellent. For $\alpha<0.5$, the asymptotic value of   $f_+^*$ increases as $p$ increases (see Fig. \ref{trajectory for q=2,3,infty}(a)), while for $\alpha>0.5$, the opposite happens (see Fig. \ref{trajectory for q=2,3,infty}(b)). Moreover, the relaxation towards the fixed point is considerably slower for $\alpha=0.1$ than for $\alpha=0.9$, indicating that for higher contrarian fractions the system reaches its steady state much faster.

At the fixed point $f_+ = f_+^*$, $\left( \frac{df_+}{dt}\right)_{f_+=f_+^*} = A{f_+^*}^2+Bf_+^*+\alpha = 0$ leads to a quadratic equation in $f_+^*$, whose two solution are denoted as $r_1,r_2$ (see Appendix B). Of these two, only $r_1$ is an lies
in the range [0,1].

\begin{figure*}
    \centering
    \includegraphics[width=\textwidth]{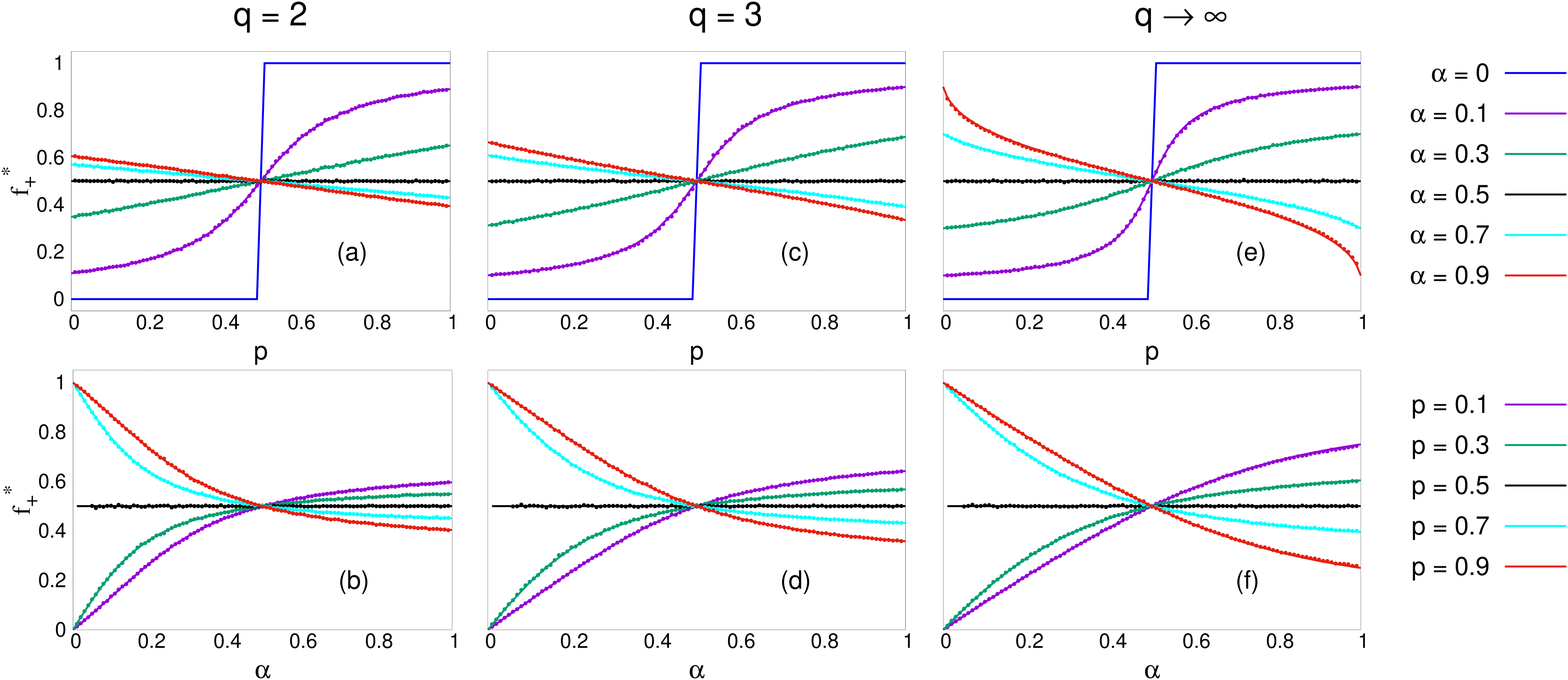}
    \caption{Variations of the stable fixed point $f_+^*$ as functions of $p$ and $\alpha$, for several values of $q$. Panels (a), (c) and (e) show the variations for $q=2$, $q=3$ and $q\rightarrow\infty$ respectively, as a function of $p$ for some typical values of $\alpha$. On the other hand, panels (b), (d) and (f) show the variations for $q=2$, $q=3$ and $q\rightarrow\infty$ respectively, as a function of $\alpha$ for some typical values of $p$.
    Simulated results are shown by solid circles and analytical results are shown by solid lines. Analytical results for $q\rightarrow\infty$ are compared with simulated results for $q=50$ and the agreement seems excellent. Simulations were performed using a system size of $N=2^{10}$ averaging over $10^2$ configurations.}
    \label{fp_p_alpha}
\end{figure*}

The stable fixed point $r_1$ implies that for any nonzero $\alpha$  trivial fixed points $f_+^*=0,1$ do not exist, instead, a single nontrivial stable fixed point appears between $0$ and $1$, whose value depends on $p$. For $p=0.5$, there exists a single fixed point $f_+^*=0.5$ independent of any non-zero value of $\alpha$. Therefore if we plot $f_+^*(p)$ for different values of $\alpha$, all the curves intersect at the point $(p,f_+^*)=(0.5,0.5)$, as shown in Fig. \ref{fp_p_alpha}(a). Similarly for $\alpha=0.5$, the only fixed point is also $f_+^*=0.5$, independent of $p$. Hence if we plot $f_+^*(\alpha)$ for different $p$, all the curves meet at the point $(\alpha,f_+^*)=(0.5,0.5)$, as shown in Fig. \ref{fp_p_alpha}(b).

From Fig. \ref{fp_p_alpha}(a), we observe that for $\alpha=0$ the system has two trivial fixed points, $f_+^*=0$ and $1$. Their stability depends on the values of $p$. For $p<0.5$, $f_+^*=0$ is stable and $f_+^*=1$ is unstable, leading to negative consensus, while for $p>0.5$ their stabilities are exchanged, leading to positive consensus. This indicates an exchange of stability bifurcation at the critical value of the parameter $p=0.5$ \cite{mullick2025social}. When $\alpha$ is increased from zero, these trivial consensus states disappear, and the system indeed reaches intermediate fixed points. For small $\alpha$, 
$f_+^*$ increases with $p$ in a nonlinear fashion before saturating at a value less than unity. For larger $\alpha<0.5$, the growth of $f_+^*$ with $p$ appears to become  linear with positive slope. At $\alpha=0.5$, the fixed point is always at $f_+^*=0.5$ independent of $p$. For $\alpha>0.5$, the variation of $f_+^*$ with $p$ seems to remain linear, but the slope becomes negative, so that $f_+^*$ decreases with increasing $p$.

In Fig. \ref{fp_p_alpha}(b), the variation of $f_+^*$ with $\alpha$ is shown for fixed values of $p$.  At $p=0.5$, the fixed point remains exactly at $f_+^*=0.5$ for all nonzero $\alpha$.  This overall behavior is consistent with the intrinsic symmetry of the dynamical equation for $f_+$ under the transformation $p\to 1-p$ and $f_+\to 1-f_+$, which ensures the dynamics for $p<0.5$ and $p>0.5$ are related in a symmetric way. However, no such symmetry for $\alpha$ exists which can be checked from Eq. \ref{simplified master equation}.

\subsubsection{q = 3 case}
Now for $q=3$, the master equation for $f_+(t)$, i.e. Eq. (\ref{simplified master equation}), reduces to
\begin{equation}
\label{simplified master equation for q=3}
\begin{split}
\frac{df_+}{dt} = \left[\frac{3p}{2-p}-\frac{6p}{1+p}-\frac{6\alpha p}{2-p}+\frac{12\alpha p}{1+p}+1-2\alpha\right]f_+^{3}\\
+\left[\frac{6p}{1+p}-\frac{6p}{2-p}+\frac{12\alpha p}{2-p}-\frac{12\alpha p}{1+p}\right]f_+^{2}
\\ 
+\left[\frac{3p}{2-p}-\frac{6\alpha p}{2-p}-1\right]f_+ + \alpha
\end{split}
\end{equation}

Unlike the case $q=2$, where a closed form solution of $f_+(t)$ is available, for $q=3$ this cubic differential equation does not admit a closed expression for the trajectory. We therefore solve Eq. (\ref{simplified master equation for q=3}) numerically. The numerical solution of Eq. (\ref{simplified master equation for q=3}) are shown in Fig. \ref{trajectory for q=2,3,infty}(c) for $\alpha=0.1$ and in Fig. \ref{trajectory for q=2,3,infty}(d) for $\alpha=0.9$. For a given value of $p$, the trajectory $f_+(t)$ approaches the fixed point much more slowly for $\alpha=0.1$, indicating a larger relaxation time, whereas for $\alpha=0.9$ the relaxation is significantly faster.


Eq. (\ref{simplified master equation for q=3}) at fixed point $f_+^*$, is a cubic equation in $f_{+}^*$ such that one can expect three fixed points.
However, only one stable fixed point lies between $[0,1]$ which depends on $p$. The dependency of $f_+^*$ on $p$ for different values of $\alpha$ and the dependency of $f_+^*$ on $\alpha$ for different values of $p$ is quite similar to the dependency of $f_+^*$ for $q=2$ case. And in this case also, all the curves $f_+^*(p)$ for different values of $\alpha$ intersect at $p=0.5,f_+^*=0.5$ [Fig. \ref{fp_p_alpha}(c)]. And also, the curve $f_+^*(\alpha)$ for $p=0.5$ is a straight line parallel to $\alpha$ axis [Fig. \ref{fp_p_alpha}(d)].

Although the results for both $q=2,3$ have been shown for the initial condition $f_0=0$, since there is only one stable fixed point, $f_+^*$ values are independent of the initial values as confirmed both by the analytical equations and numerical simulations. 


\subsection{Larger $q$ values : Simplified mean field theory}
\label{simpmf}
So far we considered mean field theory assuming $N\to \infty$ but $q$ finite. However, as the first term in Eq. (\ref{simplified master equation}) becomes cumbersome for large $q$, we make an additional approximation and replace $n$ by its average value which is $qf_+$ in the sum in Eq. (\ref{simplified master equation}). We call this the simplified mean field theory and the above approximation is valid for $q\to \infty$.

Using $n=qf_+$ in Eqs (\ref{negative to positive rate}) and (\ref{positive to negative rate}), the transition rates $\omega$ between a positive (+) state and a negative (-) state are obtained as
\begin{widetext}
\begin{equation}
\label{rate from negative to positive}
\omega_{-\to +} = \alpha \left[(1-f_+)^{q}+ \frac{(1-f_+)(1-p)[1-f_+^q-(1-f_+)^q]}{f_+p+(1-f_+)(1-p)}\right]
+(1-\alpha)\left[f_+^q +\frac{f_+p [1-f_+^q-(1-f_+)^q]}{f_+p+(1-f_+)(1-p)} \right]
\end{equation}
\begin{equation}
\label{rate from positive to negative}
\omega_{+\to -} = (1-\alpha)\left[(1-f_+)^q+ \frac{(1-f_+)(1-p)[1-f_+^q-(1-f_+)^q]}{f_+p+(1-f_+)(1-p)}\right]
+\alpha\left[f_+^q+\frac{f_+p[1-f_+^q-(1-f_+)^q]}{f_+p+(1-f_+)(1-p)}\right]
\end{equation}
Now the rate equation simplifies to
\begin{equation}
\label{SMF master equation}
\begin{split}
\frac{df_+}{dt} = (1-\alpha)\left[(1-f_+)f_+^q-f_+(1-f_+)^q+\frac{f_+(1-f_+)[1-f_+^q-(1-f_+)^q](2p-1)}{f_+p+(1-f_+)(1-p)}\right]\\[0.3cm]
+\alpha\left[(1-f_+)^{q+1}-(f_+)^{q+1}+\frac{[1-f_+^q-(1-f_+)^q][(1-f_+)^2(1-p)-f_+^{2}p]}{f_+p+(1-f_+)(1-p)}\right]
\end{split}    
\end{equation}
\end{widetext}
For $q\to \infty$ limit, the Eq. \ref{SMF master equation} becomes
\begin{equation}
\label{simplified mean field rate equation for q = infinity}
\frac{df_+}{dt} = \frac{(1-2p)f_+^2+(2p-\alpha -1)f_++\alpha - \alpha p}{f_+p+ (1-f_+)(1-p)}
\end{equation}
If we put $p=0.5$ in the above equation, it simplifies to $\frac{df_+}{dt} = \alpha(1-2f_+)$ which is identical to the original mean field Eq. (\ref{simplified rate equation involving <n>}) at $p=0.5$. For any nonzero $\alpha$ one can easily write down the solution as
\begin{equation}
\label{solution for p =0.5 and q tends to infinity}
f_+(t)\big|_{p=0.5} = \frac{1}{2}-\left(\frac{1}{2}-f_0\right)\exp(-2\alpha t).
\end{equation}

The trajectory of $f_+$ as a function of time $t$ is plotted for contrarian probability $\alpha = 0.1$ [Fig. \ref{trajectory for q=2,3,infty}(e)] and for $\alpha = 0.9$ 
[Fig. \ref{trajectory for q=2,3,infty}(f)].


The  fixed point $f_+ = f_+^*$ is given by,  
\begin{equation}
\label{fixed point calculation for q = infinity}
 (1-2p)(f_+^*)^2+(2p-\alpha -1)f_+^* +\alpha - \alpha p = 0 ,  
\end{equation}
provided  $f_+^*p+(1-f_+^*)(1-p)$ is non zero for a particular $f_+^*$ and $p$.  All the exceptional cases are discussed in the Appendix C. Once again, linear stability analysis shows the existence of only one stable fixed point  $f_+^*$ in the range [0,1] as shown in Appendix C.

 The behavior of $f_+^*$ with $p$ in this case for different values of $\alpha$ is plotted in Fig. \ref{fp_p_alpha}(e) and also the behavior of $f_+^*$ with $\alpha$ for different values of $p$ is plotted in Fig. \ref{fp_p_alpha}(f).


\subsection{Dependence on $q$ and various phases}

\begin{figure}[h!]
    \centering
    \includegraphics[width=\linewidth]{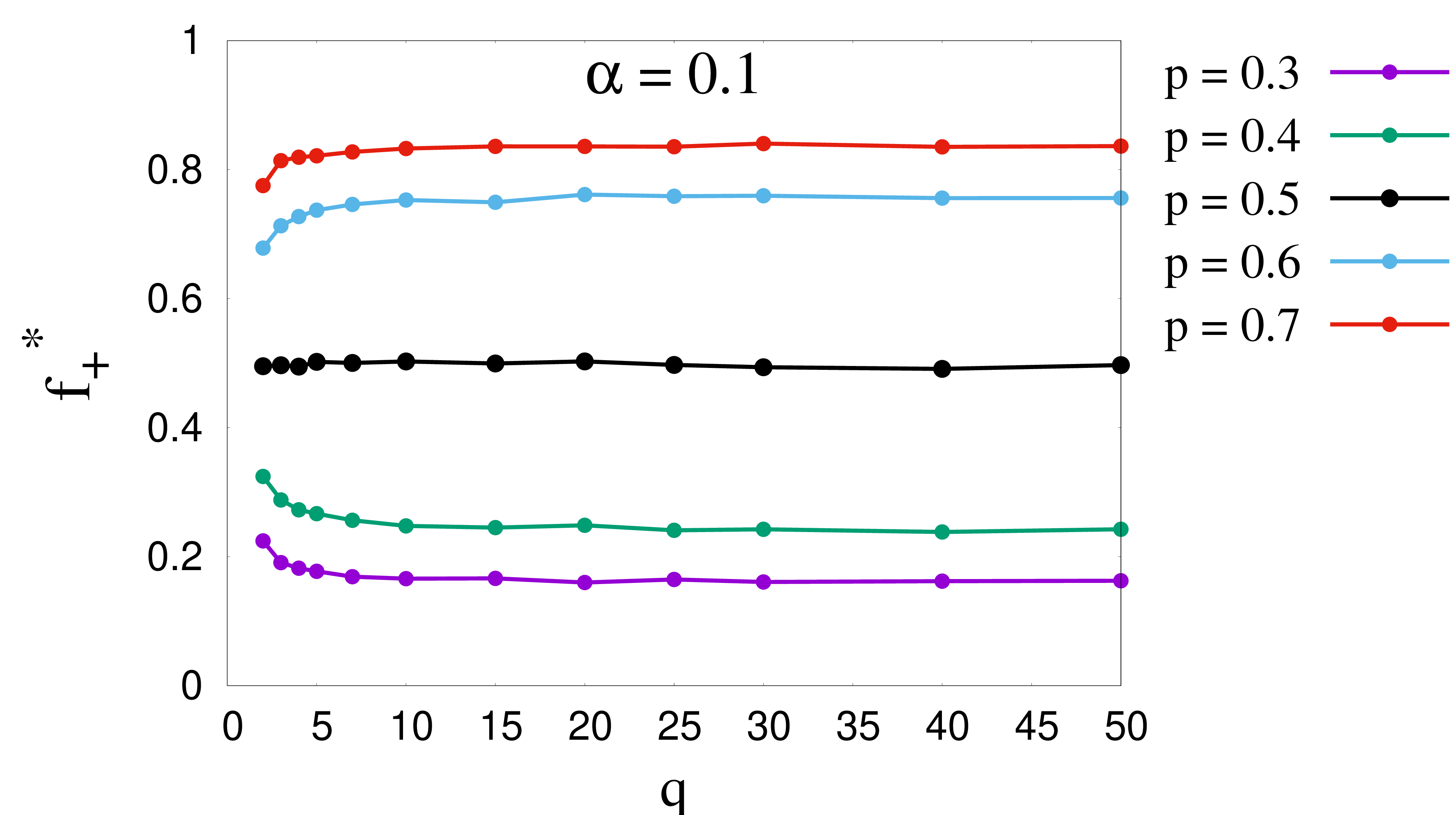}
    \caption{Variation of stable fixed points $f_+^*$ as a function of $q$ for several values of $p$ and $\alpha=0.1$. $f_+^*$ shows a weak dependence on $q$ for small $q$, but becomes independent of $q$ as $q$ increases. These results have been obtained using numerical simulations. The solid lines are guides to the eye.}
    \label{fig:fp_vs_q}
\end{figure}

It is important to study the $q$-dependence of our results. In Fig. \ref{fig:fp_vs_q} we show the variation of the stable fixed points $f_+^*$ as a function of $q$ for several values of $p$ and a fixed value of $\alpha$, the results are not sensitive to the particular value of $\alpha$. For all $p$, except $p=1/2$, we see that $f_+^*$ shows some dependence on $q$ for small values of $q$. However, at larger values of $q$, $f_+^*$ becomes saturated (see Eq. (\ref{q independence of dynmaical equation}) of Appendix D). This behavior was also observed in \cite{mullick2025social} for $\alpha =0$. For $p=1/2$, we have already established that the stable fixed point remains $1/2$ for any $q$, which is also shown in Fig. \ref{fig:fp_vs_q}. 

\begin{figure*}
    \centering
    \includegraphics[width=\textwidth]{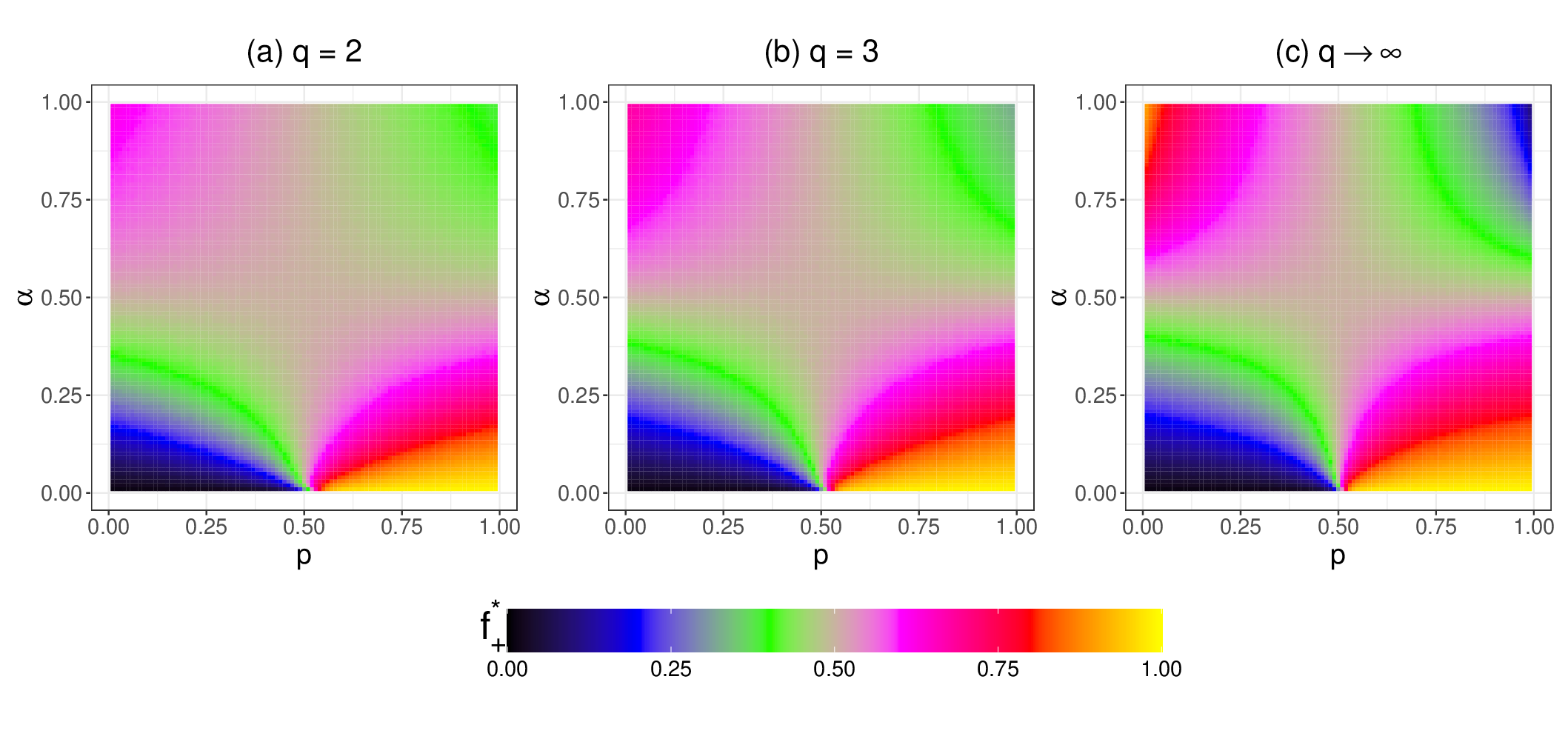}
    \caption{Heatmaps of the fixed point density $f_+^{*}$ in the $(p,\alpha)$ plane for (a) $q=2$, (b) $q=3$ and (c) $q\to \infty$ are shown where $0 < p,~\alpha < 1$. Color scale denotes $f_+^*$ from low (black) to high (yellow). For $\alpha<0.5$, $f_+^*$ increases monotonically with $p$, and the saturation value attained at large $p$ decreases as $\alpha$ is raised from $0$ to $0.5$. For $\alpha>0.5$, $f_+^*$ decreases monotonically with $p$. This trend is qualitatively similar for $q=2$ and $q=3$ while for $q\to \infty$ the transition between low and high $f_+^*$ becomes sharper and more step-like.}
    \label{fig:fp_heatmaps}
\end{figure*}

To comprehensively demonstrate the $q$-dependence of the fixed points, we then plot the variation of $f_+^*$ as a function of $p$ and $\alpha$, i.e., a heatmap, for several values of $q$. The heatmaps are shown in Fig. \ref{fig:fp_heatmaps}, from which we can clearly see that $f_+^*$ has an interesting dependence on $p$ and $\alpha$, which again also depends on the value of $q$. As already demonstrated in Figures 2, 4, and 6, for $p, \alpha<0.5$ and $p, \alpha>0.5$ all the fixed points are $<0.5$, but they are $>0.5$ when $p<0.5$, $\alpha>0.5$ and $p>0.5$, $\alpha<0.5$. The value of $f_+^*$ being either greater than $0.5$ or less than $0.5$ indicates that the final state of the system has one of the opinions as the majority, even in the presence of contrarians. However, (i) for $\alpha=0.5$, i.e., when half the population are contrarians, and (ii) for $p=0.5$, i.e., when the population is unbiased, we get $f_+^*=0.5$, indicating a situation where both the opinions have equal fractions in the system, termed as ``hung election" in \cite{galam2004contrarian}.


\begin{figure}[h!]
    \centering
    \includegraphics[width=\linewidth]{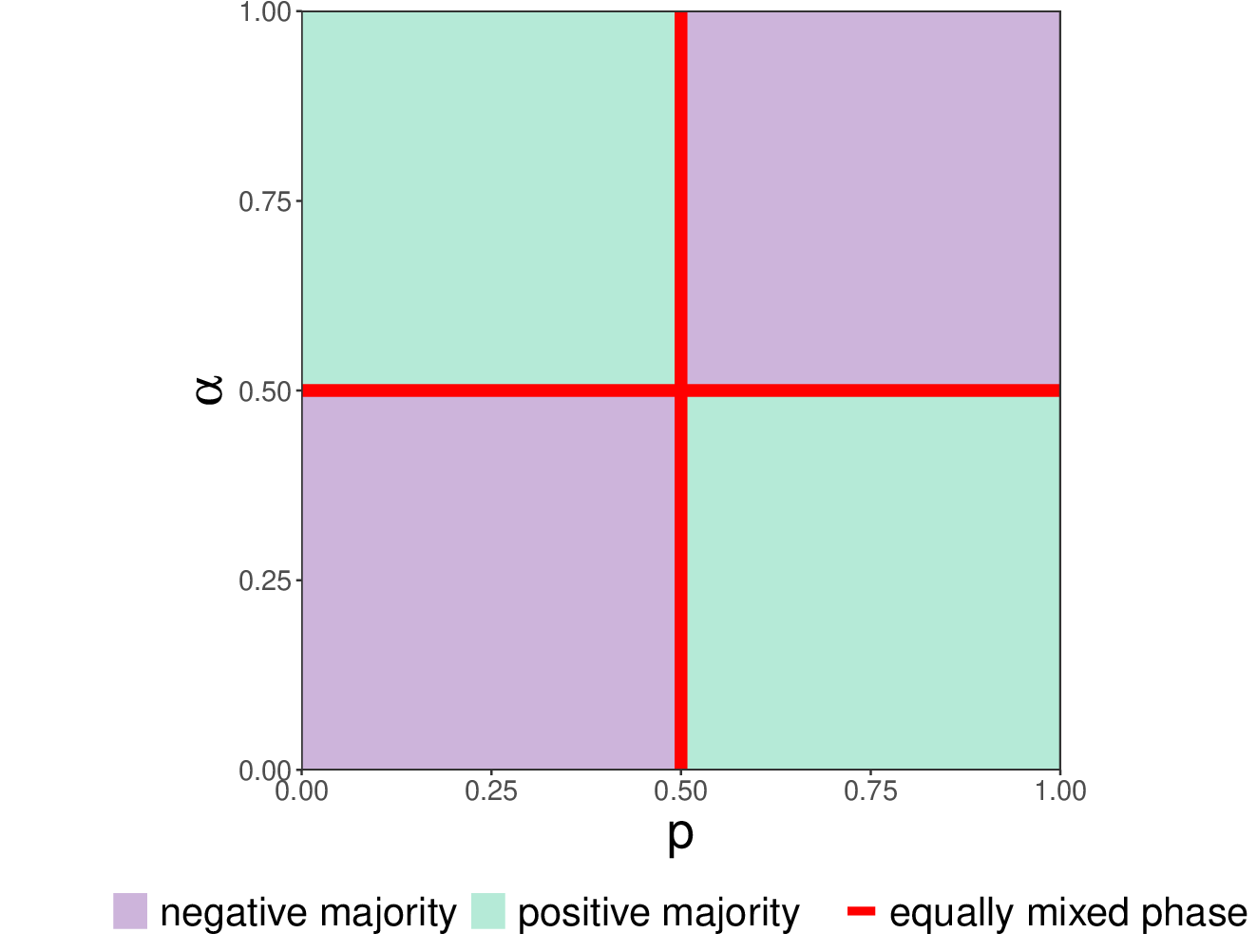}
    \caption{Phase diagram  in the $(p,\alpha)$ parameter space shows distinct dynamical regimes: the negative majority phase (purple) and the positive majority phase (green), separated by the central red lines corresponding to the equally mixed phase. Existence of these phases is independent of the value of $q$.}
    \label{fig:phases}
\end{figure}

Clearly there exists several phases in our system depending on the values of $p$ and $\alpha$. In Fig. \ref{fig:phases} we show a schematic diagram representing the phases in the $(p,\alpha)$ plane. The phase with equally mixed opinions are obtained only along the lines $\alpha=0.5$ and $p=0.5$. For any other values of $\alpha$ and $p$, a state with a clear majority is obtained. When $\alpha<0.5$, the majority in the final state depends on the value of $p$. If $p<0.5$ ($p>0.5$), i.e., if agents with negative (positive) opinion have higher influential power, the system goes to a state with negative (positive) majority. As soon as $\alpha$ increases beyond $0.5$ this outcome is reversed, clearly because of the higher fraction of contrarians who opposes any social influence. Interestingly, even for higher fractions of contrarians the system does not get stuck in a phase with equal fractions of the two opinions, as shown in earlier studies \cite{nyczka2012phase,nyczka2013anticonformity,anugraha2025nonlinear}. This is certainly a consequence of the bias $p$, in the form of individual influential powers, that is present in our model.

\subsection{Timescales for different $q$}

\subsubsection{$q=2$}


For $q=2$ we have an explicit closed form solution for $f_+(t)$ given in Eq. (\ref{Exact expression of f_+ for q=2}). However, extracting a well defined relaxation timescale directly from this solution is not straightforward. At the same time $f_+(t)$ (see Fig. \ref{trajectory for q=2,3,infty}), can be  fitted quite accurately to the functional form $f_+(t) = f_+^*(p)(1-\exp(-t/\tau(p)))$ starting from the initial condition $f_0=0$. Consistently, for the initial condition $f_0\neq0$,  the trajectory is accurately fitted by $f_+(t)=f_+^*+(f_0-f_+^*)\exp(-t/\tau)$. 
We therefore define $\tau$ as the characteristic global relaxation timescale which characterizes how fast the system relaxes to stable fixed point when it starts from some initial point far from fixed point. $\tau$ can be obtained by solving $f_+(\tau)=f_+^*+(f_0-f_+^*)\exp(-1)$. A closed form expression (see Eq. (\ref{tau for alpha,p neq 0.5}) of Appendix B) for $\tau$ is given by 


\begin{equation}
\label{Global timescale for q = 2 for general f0}
\tau = \frac{1}{\sqrt{\Delta}}\ln \left[\frac{2Af_0+B+\sqrt{\Delta}-2e\sqrt{\Delta}}{2Af_0+B-\sqrt{\Delta}}\right].
\end{equation}
Here, the quantities $A$, $B$ and $\Delta$ appearing in the expression of $\tau$ are already defined in Eq. (\ref{Simplified master equation for q=2}). 



\begin{figure}[h]
\includegraphics[width=\linewidth]{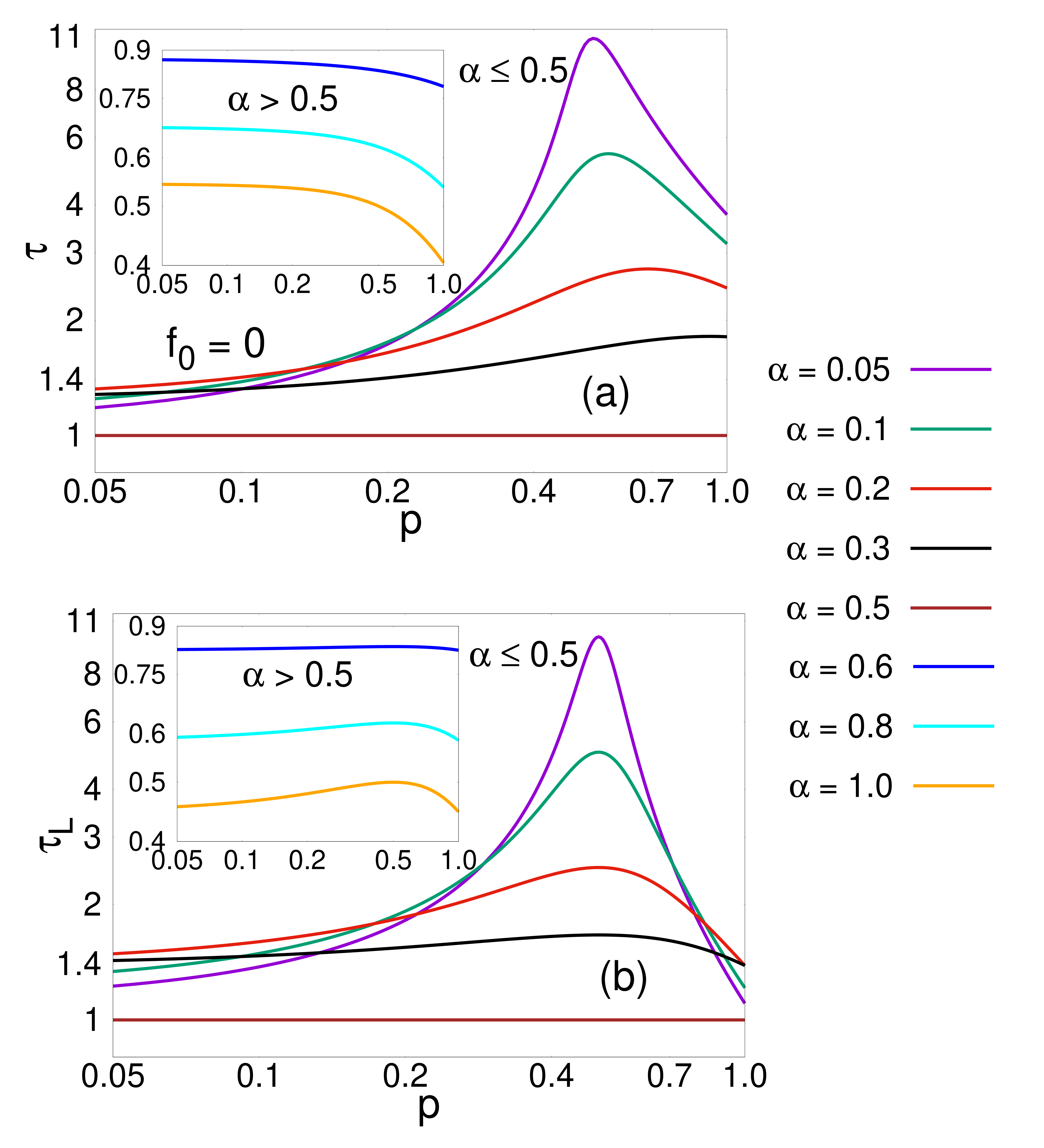} 
\caption{(a) Plot of global time scale $\tau$ in log scale
as a function of the positive influential power $p$
is shown for different values of $\alpha$ where $0< \alpha \leq 0.5$. Inset shows the variation of $\tau$ with $p$ for different values of $\alpha$ where $0.5<\alpha \leq  1$. Plots are for initial condition $f_0=0$ (b) Variation of local timescale $\tau_L$  in log scale as a function of positive influential power $p$ is shown for different values of $\alpha$ where $0< \alpha \leq 0.5$. Inset shows the variation of $\tau_L$ with $p$ for different values of $\alpha$ where $0.5<\alpha \leq  1$. All the plots are for $q=2$.}
\label{consensus time for q = 2}
\end{figure}

Eq. (\ref{Global timescale for q = 2 for general f0}) for the time scale $\tau$ for any $f_0$ is valid for all $\alpha$ and $p$, except for $\alpha = 0.5$ and $p=0.5$. For $\alpha = 0.5$, $f_+^* = 0.5$ for all $p$, therefore we obtain a $p$ independent timescale $\tau_{\alpha = 0.5} = 1$ and for $p=0.5$ we have an $\alpha$ dependent timescale $\tau_{p=0.5} = {1}/{2\alpha}$ for all non-zero alpha. Remarkably these expressions turn out to be independent of the initial condition $f_0$ . The variation of $\tau$ with $p$ for different values of $\alpha$ for initial condition $f_0=0$ is shown in Fig. \ref{consensus time for q = 2}(a).


In addition, a local relaxation timescale $\tau_{L}=1/\sqrt{\Delta}$ can be obtained from the stability analysis of the fixed point, which characterizes the exponential decay of small perturbations close to the stable fixed point (see Appendix B). The variation of $\tau_{L}$ with $p$ for different values of $\alpha$ is shown in Fig. \ref{consensus time for q = 2}(b). The data for local and global timescales are qualitatively very similar. 

\begin{figure}[h]
\includegraphics[width=\linewidth]{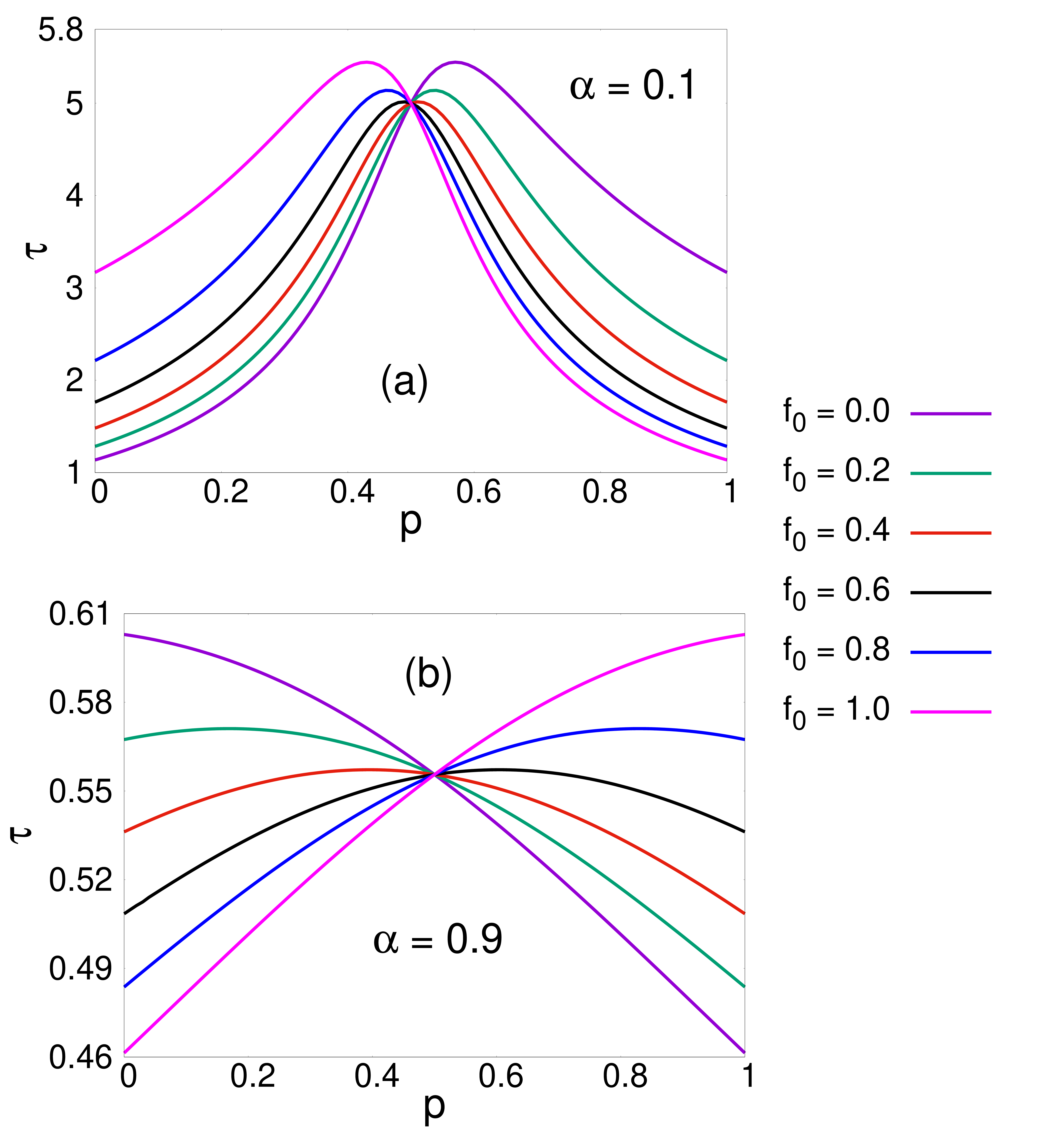} 
\caption{Plots of the global relaxation time scale $\tau$ (which is the time at which the deviation of $f_+(t)$ from its asymptotic value $f_+(t\to \infty)$ has decayed to $1/e$ of its initial deviation) as a function of the positive influential power $p$ are shown for different initial values of $f_0$  at (a) $\alpha=0.1$ (b)  $\alpha=0.9$. All the plots are for $q=2$.}
\label{timescale_p_for_different_f0}
\end{figure}

The global relaxation timescale $\tau$ depends explicitly on the initial condition $f_0$, whereas the local timescale $\tau_{L}$ does not, since it follows directly from the stability analysis of the fixed point. The local timescale is thus universal in the sense that it is governed solely by the fixed point, which determines the steady state behavior of the system. In Fig. \ref{timescale_p_for_different_f0}, we plot the global timescale $\tau$ as a function of $p$ for different initial conditions at two fixed values of $\alpha = 0.1,0.9$.

\subsubsection{$q=3$ and $q\to \infty$}

\begin{figure}[h]
\includegraphics[width=\linewidth]{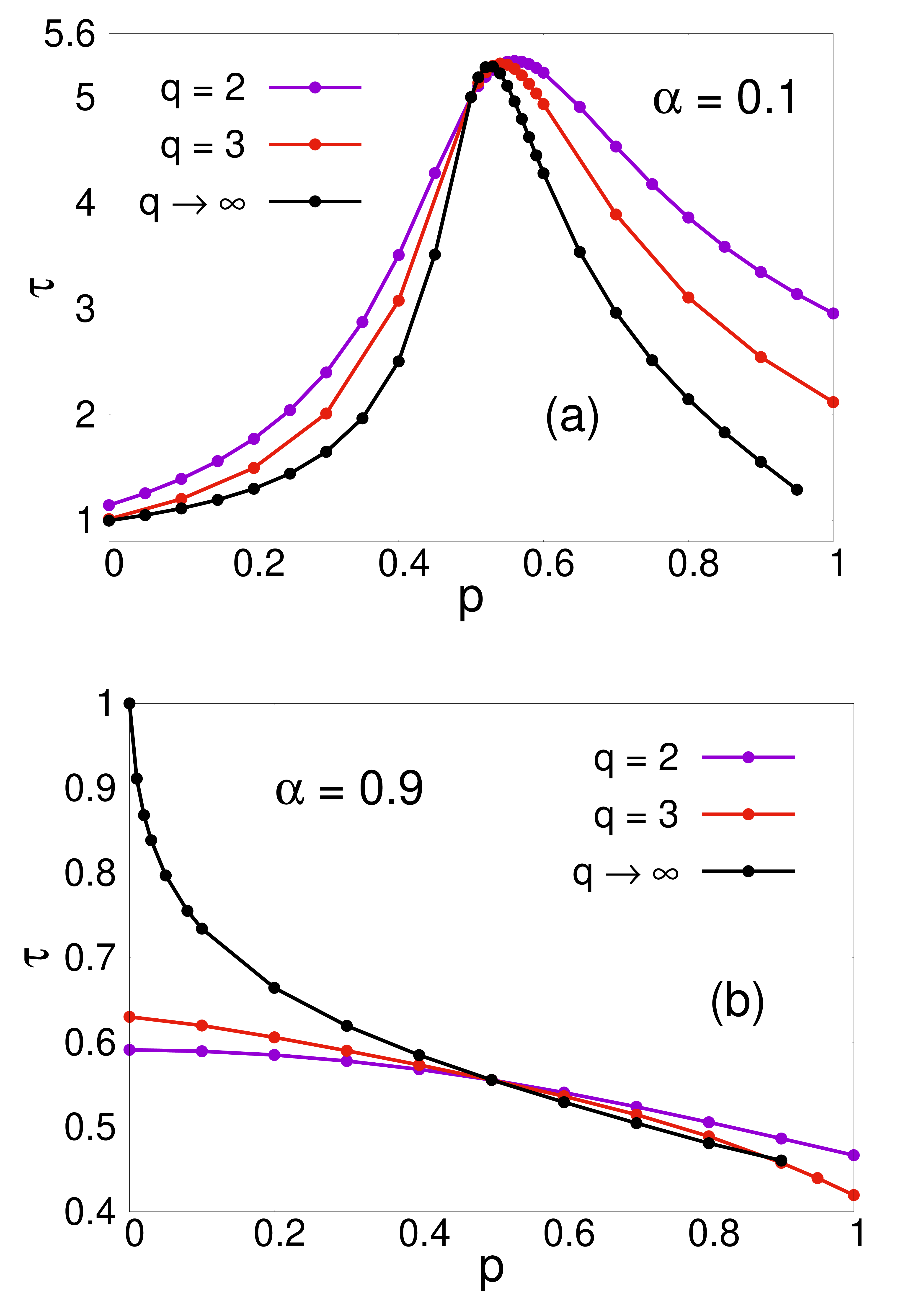} 
\caption{(a) Plot of the global time scale $\tau$  as a function of the positive influential power $p$ is shown for different values of $q$ at $\alpha=0.1$. Peak of $\tau$ is at $p_c= 0.56,0.54,0.53$ for $q=2,3$ and $q\to \infty$ respectively. (b) Variation of the global time scale $\tau$ with $p$ for different values of $q$ at $\alpha = 0.9$ is shown. All plots are for $f_0=0$.}
\label{consensus time}
\end{figure}

Unlike the case $q=2$, where closed form expressions exist for the solution $f_+(t)$ and both local and global timescales, no such closed forms are available for $q=3$. For $q\to \infty$,  the complete solution $f_+(t)$ is not analytically accessible, and therefore a closed form expression for the global timescale does not exist. However, a closed form expression for the local timescale can still be obtained from the stability analysis (see Appendix C). For $q=3$ and $q\to \infty$, one can estimate the global relaxation timescale $\tau(p)$ by fitting the trajectory $f_+(t)$ (starting from $f_0=0)$ with the functional form $f_+^*(p)[1-\exp{(-{t}/{\tau(p)})}]$ for a fixed $\alpha$. The resulting variation of $\tau(p)$ with $p$ is shown in Fig. \ref{consensus time}.
For $\alpha=0.1$ (Fig. \ref{consensus time}(a)), $\tau(p)$ displays a non-monotonic dependence on $p$: it increases with $p$, attains a maxima close to $p=0.5$ and then decreases sharply. As $q$ increases, this maxima shifts slightly to the left, and the overall relaxation becomes faster, with all curves $\tau(p)$ for different values of $q$ intersecting at $p=0.5$. This is consistent with the unique symmetric fixed point $f_+^*=0.5$ for $p=0.5$ that exists for any $\alpha>0$. In contrast, for $\alpha=0.9$ (Fig. \ref{consensus time}(b)), the peak near $p=0.5$ is entirely absent. For $\alpha>0.5$, $\tau(p)$ decreases monotonically across the full range of $p$. In this regime, the relaxation is significantly faster compared to the case $\alpha<0.5$ and the dependence of global timescale $\tau$ on $q$ is much weaker, resulting nearly overlapping $\tau(p)$ curves.

For $q\to \infty$, we also define a local relaxation timescale which characterizes how fast the system relaxes to a fixed point when it starts from a small neighborhood of that point. This timescale admits a compact closed form (see Eq. (\ref{local timescale for q tends to infinity}) of Appendix C) ,
\begin{equation}
\label{tau_L_qinfty}
  \tau_L = \frac{1-\alpha+\sqrt{\Delta^{\prime}}} {2\sqrt{\Delta^{\prime}}} 
\end{equation}
with discriminant $\Delta^{\prime} = (1-\alpha)^2 + 4p(1-p)(2\alpha-1)$

This expression is valid for all $(\alpha,p)$ except certain special cases for which discriminant $\Delta^{\prime}=0$. In fact, $\Delta^{\prime}=0$ occurs only for the three parameter combinations $(\alpha,p)=(0,1/2),(1,0),(1,1)$. At $(\alpha,p)=(1,0)$ and $(1,1)$, the limiting value of the timescale is $\tau_L=1/2$.

\begin{figure}[h]
\centering
\includegraphics[width= \linewidth]{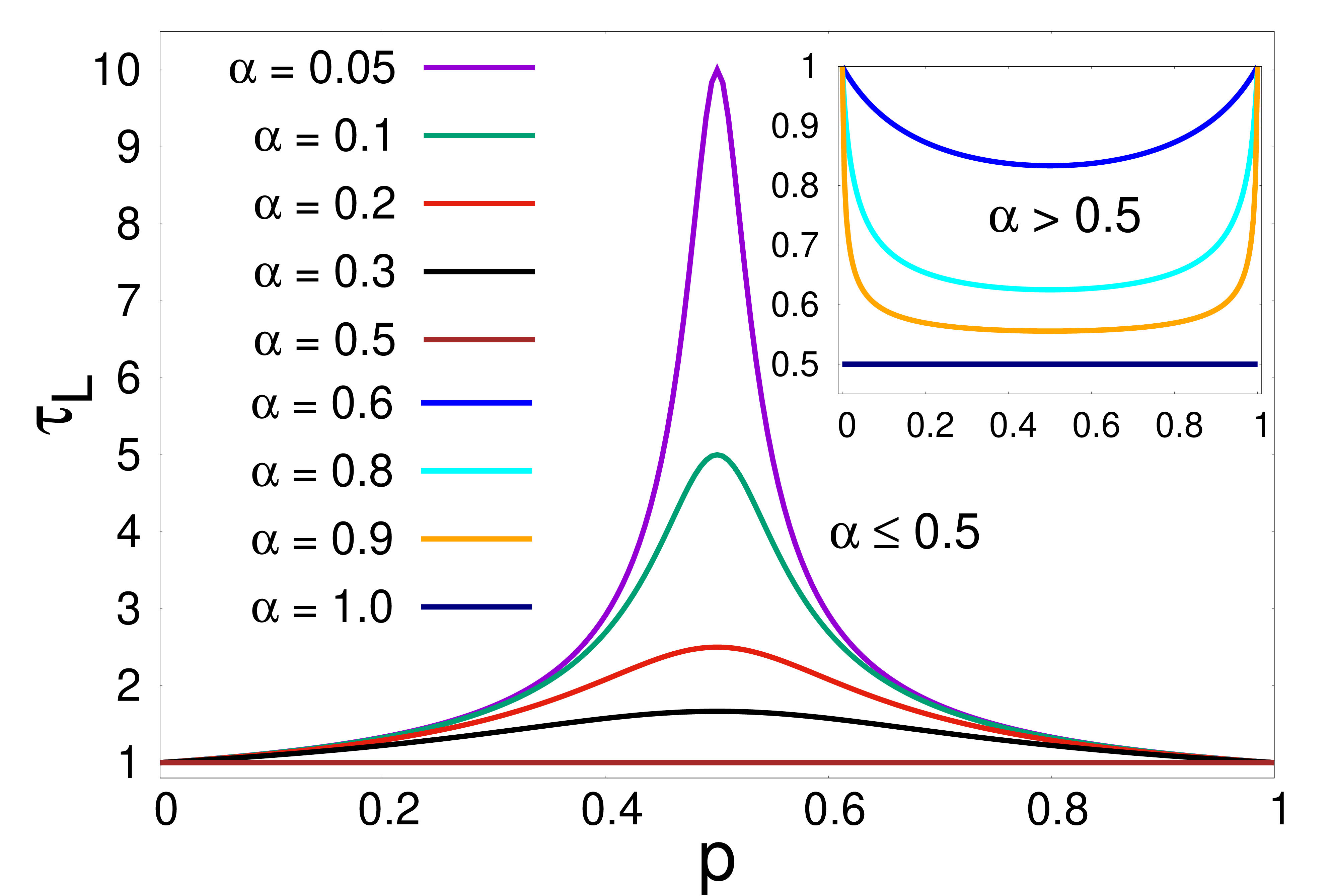} 
\caption{ Variation of the local timescale $\tau_L$ as a function of positive influential power $p$ is shown for different values of $\alpha$ where $0< \alpha \leq 0.5$. Inset shows the variation of $\tau_L$ with $p$ for different values of $\alpha$ where $0.5<\alpha \leq  1$. Plots are for $q\to \infty$.}
\label{local_timescale_qinfty}
\end{figure}

A further exception arises along the line $p=1/2$ for all non-zero $\alpha$. In this case the dynamical equation for $f_+$ becomes strictly linear, with a single fixed point at $f_+^*=1/2$. The correct result for the local timescale in this case is $\tau_L=1/2\alpha$ valid for all non-zero $\alpha$. The variation of $\tau_L(p)$ with $p$ for different $\alpha$ is shown in Fig. \ref{local_timescale_qinfty}, and the detailed derivation including the treatment of these exceptional cases, is provided in the Appendix C.

\section{Summary and Conclusions}
\label{sec: conclu}

In this work, we investigated the nonlinear biased $q$-voter model in the presence of contrarians. By extending the framework introduced in Ref. \cite{mullick2025social}, where non-unanimous $q$-panels exert weighted influence through a bias parameter $p$, we considered contrarians who oppose both unanimous and non-unanimous influences. This modification leads to qualitatively distinct steady-state behavior compared to earlier studies of anti-conformist contrarians.

Our analysis, combining mean-field theory with Monte Carlo simulations of an agent based model (which show excellent agreement), manifests that a mixed state with equal support for both opinions arises only under two special conditions: when the contrarian fraction is exactly $\alpha=1/2$ or when the bias is symmetric $(p=1/2)$. In all other cases, the system evolves towards polarized steady states characterized by a clear majority, even when the contrarian fraction is high. This result sharply contrasts with previous works, where sufficiently large contrarian fractions suppressed majority formation and produced ``hung election scenario''.

We also found a symmetry under the transformation $p\rightarrow 1-p$ and $f_+\rightarrow 1-f_+$, which indicates that a reversal of the bias parameter simply exchanges the roles of the two opinions without altering the macroscopic outcome for any contrarian density $\alpha$. This reflects the neutrality of the model with respect to the choice of the dominant opinion.

The phase diagram in the $(p,\alpha)$ plane clearly demarcates regions of positive and negative majority, separated only by the lines $\alpha=1/2$ and $p=1/2$, where mixed states occur. We also observe a crossover effect that arises at $\alpha=1/2$. For $\alpha<1/2$, a bias $p>1/2$ (or $p<1/2$) produces a majority of positive (or negative) opinions, consistent with contrarians being insufficient to offset the prevailing bias. However, for $\alpha>1/2$, there is no such symmetry, as replacing $\alpha$ with $1-\alpha$ does not yield a mirrored behavior. In this regime, increasing the fraction of contrarians reduces the dominance of one opinion, driving the system closer to a balanced state with $f_+$ tending toward $1/2$. Yet, the bias $p$ ensures that the system never reaches an exact 50-50 split, leaving one opinion with a persistent marginal advantage. Importantly, such an exact balance is an exceptional and rare condition: it occurs only in a very narrow portion of the $(p,\alpha)$ parameter space, consistent with the rarity of perfectly tied outcomes in collective decision-making processes such as political elections.

We further established that the dependence of the steady states on the group size $q$ is weak. For small $q$, minor deviations are observed, but as $q$ increases the results rapidly converge and become effectively independent of $q$. We also analyzed the relaxation dynamics and derived global and local timescales, showing that the approach to steady state is significantly faster at higher contrarian fractions.

Taken together, our findings demonstrate that contrarians in the biased nonlinear $q$-voter model reinforce polarization rather than neutralize it, thereby ensuring the emergence of a clear majority across almost the entire parameter space. This robustness of majority formation, even in the presence of substantial contrarian behavior, constitutes the central result of our study. Future work may extend these ideas to structured networks or models with heterogeneous biases, which could provide further insights into the role of systematic opposition in collective decision-making.

\begin{acknowledgments}
AP would like to acknowledge University Grant Commission (UGC), Govt. of India for financial support (Student ID: 241610061476). PM acknowledges the invitation from Department of Physics, University of Calcutta. PS thanks financial support from CSIR scheme no 03/1495/23/EMR-II.
\end{acknowledgments}

\vspace{1cm}

\appendix


\section{Analytical solution of the master equation in $f_+$ for $q=2$.}
In this appendix, we analytically solve the differential equation governing the dynamics of $f_+(t)$ for the case $q=2$. Starting from the simplified form of the master equation, we rewrite the equation in an integrable form and obtain an explicit solution for $f_+(t)$ in terms of system parameters $\alpha$ and $p$.

The master Eq. (\ref{simplified master equation}) for $q=2$, reduces to
\begin{equation}
\label{simplified master equation for q=2}  
 \frac{df_+}{dt} = Af_{+}^{2} + Bf_+ + \alpha.
\end{equation}
We use the concise notation $A = 4\alpha p-2\alpha-2p+1$, $B = 2p-1-4\alpha p$ and $\Delta = B^2-4A\alpha$. A regular fixed point $f_+^*$ satisfies 
$\left(\frac{df_+}{dt}\right)_{f_+=f_+^*}=0$, which involves the discriminant $\Delta$ as $f_+^*=-(\sqrt{\Delta}+B)/2A$. Since $f_+$ is, by definition, a probability that an agent takes a positive opinion, it must remain real and satisfy $0\leq f_+\leq 1$. Consequently, for all $\alpha,p \in[0,1]$, the discriminant is strictly positive, $\Delta>0$ . By taking this into account, one can  write down the solution in closed form  as
\begin{equation}
\label{analytical solution for q = 2}
 t = \frac{1}{\sqrt{\Delta}}\ln\left [\frac{2Af_++B-\sqrt{\Delta}}{2Af_++B+\sqrt{\Delta}}\right] + \lambda .  
\end{equation}
The initial condition $f_+(t=0) = f_0$ determines the value of the constant $\lambda$ as $\lambda = -\frac{1}{\sqrt{\Delta}}\ln\left [\frac{2Af_0+B-\sqrt{\Delta}}{2Af_0+B+\sqrt{\Delta}}\right]$. By putting the value of $\lambda$ in Eq. (\ref{analytical solution for q = 2}) and then after some algebra, one can obtain a closed form solution for $f_+(t)$ as 
\begin{equation}
\label{exact expression of f_+ for q=2}
    f_+(t) = \frac{\sqrt{\Delta}}{2A}\left[\frac{1+\left(\frac{2Af_0+B-\sqrt{\Delta}}{2Af_0+B+\sqrt{\Delta}}\right)e^{\sqrt{\Delta}t}}{1-\left(\frac{(2Af_0+B-\sqrt{\Delta}}{2Af_0+B+\sqrt{\Delta}}\right)e^{\sqrt{\Delta}t}}\right]-\frac{B}{2A}.
\end{equation}
The asymptotic value of the solution $f_+(t)$  in the limit $t\to\infty$ (which is indeed the fixed point) is given as $f_+(t\to \infty) = f_+^* = -(\sqrt{\Delta}+B)/2A$.

Now, either $\alpha=0.5$ or $p=0.5$ makes $A=0$. So $f_+(t)$ is the solution of the differential 
Eq. (\ref{simplified master equation for q=2}) for all $\alpha, ~p$ except when $\alpha=0.5$ and $p=0.5$.

For $\alpha = 0.5$, Eq. (\ref{simplified master equation for q=2}) reduces to $\frac{df_+}{dt} = 1/2-f_+$. Integrating the equation yields $\ln|1/2-f_+| = -t +\lambda$.By imposing the initial condition $f_+(t=0)=f_0$  one can obtain the closed form solution for $f_+(t)$ in this case as 
\begin{equation}
\label{solution for alpha = 0.5}
 f_+(t)\big|_{\alpha=0.5} = \frac{1}{2}-(\frac{1}{2}-f_0)\exp(-t).
\end{equation}
such that $f_+\big|_{\alpha=0.5}(t\to \infty) = 1/2$ for all $p$.

For $p=0.5$, Eq. (\ref{simplified master equation for q=2}) reduces to $\frac{df_+}{dt} = \alpha(1-2f_+)$. For $\alpha\neq0$, solving the equation with $f_+(0)=f_0$ gives
\begin{equation}
\label{solution for p = 0.5}
 f_+(t)\big|_{\substack{p = 0.5 \\ \alpha \neq 0}} = \frac{1}{2}-(\frac{1}{2}-f_0)\exp(-2\alpha t) .  
\end{equation}
such that $f_+(t)\big|_{\substack{p = 0.5 \\ \alpha \neq 0}}(t\to \infty) =1/2$.


\section{Linear Stability Analysis and relaxation timescales for $q=2$}

In this appendix, we analyze the relaxation dynamics of the mean-field master equation for $q=2$. We first identify the fixed points and perform a linear stability analysis to obtain the local relaxation timescale, which characterizes the exponential decay of small perturbations around the stable fixed point. We then define a global relaxation timescale, valid for arbitrary initial conditions, and show that while the two timescales generally differ, they become identical in two special cases: (i) when the initial deviation from the stable fixed point is infinitesimal, and (ii) when the dynamics is linear $(p=0.5$ or $\alpha=0.5)$.

The mean field master Eq. (\ref{simplified master equation}) for $q=2$, reduces to Eq. \ref{simplified master equation for q=2} with the coefficients $A=(1-2\alpha)(1-2p)$, $B=2p-1-4\alpha p$ and the discriminant $\Delta = B^2-4A\alpha$. 

$\Delta$ can be written in a useful closed form obtained by algebraic simplification: $\Delta = 4[\alpha^2+(2\alpha-1)^2(p-1/2)^2]$, which makes $\Delta\geq0$ always. Note that for $p=0.5$ and $\alpha=0$, $\Delta=0$ which is the mean field voter model.

The fixed points satisfy $g(f_+^*)=0$. The two roots are $r_{1,2}=(-B\mp \sqrt{\Delta})/2A$ with $A\neq 0$. Derivatives of $g(f_+)$ at the fixed points satisfy $g^{\prime}(r_1)=2Ar_1+B = -\sqrt{\Delta}<0$ and $g^{\prime}(r_2) = \sqrt{\Delta}>0$. 
Therefore  $f_+^*=r_1$ is a stable fixed point and $r_2$ by a unstable fixed point. The root separation is $s=r_1-r_2 = -\sqrt{\Delta}/A$.

By putting $f_+(t)=r_1+\delta(t)$ and assuming $|\delta|\ll 1$, Taylor expansion of $g(f_+)$ about $r_1$ gives
\begin{equation}
 g(r_1+\delta) = g(r_1)+g^{\prime}(r_1)\delta+\mathcal{O}(\delta^2) = g^{\prime}(r_1)\delta+\mathcal{O}(\delta^2),   
\end{equation}
since $g(r_1) = 0$. Neglecting terms of $\mathcal{O}(\delta^2)$, to leading order we obtain
\begin{equation}
    \frac{d\delta}{dt} \approx g^{\prime}(r_1)\delta = -\sqrt{\Delta}\delta.
\end{equation}
Hence infinitesimal deviation $\delta$ decays exponentially as $\delta(t) = \delta(0)\exp(-\sqrt{\Delta}t)$, with a local relaxation timescale $\tau_{L} = 1/\sqrt{\Delta}$ that characterizes how fast the system relaxes to its stable fixed point $r_1$ when starting within a small neighborhood of that point.

We call the timescale $\tau_{L}$ as local because it relies on truncating $\mathcal{O}(\delta^2)$ terms, so it is valid when the initial deviation $\delta(0)$ is small enough so that the nonlinear terms remain negligible along the trajectory. In particular, $\tau_{L}$ is independent of the initial condition $f_0$. 

Next, we define another characteristic time scale $\tau$ at which the deviation of $f_+(t)$ from its asymptotic value $f_+(t\to \infty)=r_1$ has decayed to $\frac{1}{e}$ of its initial deviation, i.e., $f_+(\tau)$ satisfies
\begin{equation}
 \label{time scale definition}
  f_+(\tau)-r_1 = \frac{1}{e}(f_0-r_1).  
\end{equation}
Taking the closed form solution $f_+(t)$ for $\alpha,~p\neq0.5$ [Eq. (\ref{exact expression of f_+ for q=2})], one can  solve Eq. (\ref{time scale definition}) and the solution yields a closed form expression for the global relaxation time scale $\tau$ as 
\begin{equation}
\label{tau for alpha,p neq 0.5}
\tau\big|_{\substack{p \neq 0.5 \\ \alpha \neq 0.5}} = \frac{1}{\sqrt{\Delta}}\ln \left[\frac{2Af_0+B+\sqrt{\Delta}-2e\sqrt{\Delta}}{2Af_0+B-\sqrt{\Delta}}\right].
\end{equation}
For $\alpha=0.5$ and $p=0.5$, one can solve the Eq. (\ref{time scale definition}) by taking the closed form solution $f_+(t)$ from the Eq. (\ref{solution for alpha = 0.5}),(\ref{solution for p = 0.5}) respectively and the solution yields
\begin{equation}
\label{tau for alpha,p = 0.5}
\tau\big|_{\alpha=0.5} = 1 ; \tau\big|_{\substack{p = 0.5 \\ \alpha \neq 0}} = \frac{1}{2\alpha}.
\end{equation}
Eq. (\ref{tau for alpha,p = 0.5}) is turns out to be independent of initial condition $f_0$. But, Eq. (\ref{tau for alpha,p = 0.5}) is valid for all $f_0$ except $f_0=\frac{1}{2}$ because for $f_0=\frac{1}{2}$, the definition of the time scale in Eq. (\ref{time scale definition}) is completely irrelevant.

Defining the initial deviation from the fixed point $r_1$ as $\lambda_0=f_0-r_1$, Eq. (\ref{tau for alpha,p neq 0.5}) can be written in another form as 
\begin{equation}
\label{another form of tau for alpha,p neq 0.5}
 \tau\big|_{\substack{p \neq 0.5 \\ \alpha \neq 0.5}} = \frac{1}{\sqrt{\Delta}} \ln \left[\frac{\lambda_0+es}{\lambda_0+s}\right]=\frac{1}{\sqrt{\Delta}}\left[1+\ln\left(\frac{1+\frac{\lambda_0}{es}}{1+\frac{\lambda_0}{s}}\right)\right].
\end{equation}
If $\lambda_0/s \ll 1$, i.e. we start very close to that stable fixed point $r_1$, then expansion of the logarithm of Eq. (\ref{another form of tau for alpha,p neq 0.5}) upto $\mathcal{O}[(\frac{\lambda_0}{s})^2]$ gives
\begin{equation}
 \tau\big|_{\substack{p \neq 0.5 \\ \alpha \neq 0.5}} = \frac{1}{\sqrt{\Delta}}\left[1-\left(1-\frac{1}{e}\right)\left(\frac{\lambda_0}{s}\right)+\mathcal{O}\left(\left(\frac{\lambda_0}{s}\right)^2\right)\right].
\end{equation}
 Hence $\tau \to \tau_{L}=1/\sqrt{\Delta}$ as $\lambda_0 \to 0$ i.e. the global timescale reduces to the local relaxation timescale when the initial deviation is small.

Now, either $\alpha=0.5$ or $p=0.5$ makes $A=0$, therefore Eq. (\ref{simplified master equation for q=2}) is linear in $f_+$
\begin{equation}
\label{linear equation of f+ for A = 0}
  \frac{df_+}{dt} = Bf_++\alpha \quad \text{with} \quad f_+^* = -\frac{\alpha}{B}.
\end{equation}
The solution of Eq. (\ref{linear equation of f+ for A = 0}) can be expressed as 
\begin{equation}
 f_+(t)-f_+^* = (f_0-f_+^*)\exp(-|B|t) \quad \text{with} \quad \tau = \frac{1}{|B|}.
\end{equation}
Linearizing about $f_+^*$ gives the same linear equation for $\delta(t) = f_+(t)-f_+^*$ as $\frac{d\delta}{dt}  = B\delta$ with the solution
\begin{equation}
   \delta(t) = \delta(0)\exp(-|B|t) \quad \text{with} \quad \tau_{L} = \frac{1}{|B|}.
\end{equation}
Thus when $A=0$ i.e. either $p=0.5$ or $\alpha=0.5$, the global and the local timescales are identical because the dynamical equation for $f_+$ is linear. Therefore,
\begin{equation}
 \tau\big|_{\alpha=0.5}=\tau_{L}\big|_{\alpha=0.5} = 1 ; \tau\big|_{\substack{p = 0.5 \\ \alpha \neq 0}} =\tau_{L}\big|_{\substack{p = 0.5 \\ \alpha \neq 0}} = \frac{1}{2\alpha} .  
\end{equation}

\section{Linear Stability Analysis and local relaxation timescales for $q\to \infty$}

In this Appendix, we derive the local relaxation timescale of the fixed points associated with the time evolution of $f_+$ in the $q \to \infty$ limit. Starting from the dynamical Eq. (\ref{simplified mean field rate equation for q = infinity}) for $f_+(t)$, we identify the regular fixed points and analyze their stability by linearizing around them and obtain a general expression for the local timescale $\tau_L$. Special cases, including the boundary values ($p=0,1)$ for $\alpha=1$ where degenerate fixed point arises, are treated carefully via limiting procedures to correctly capture stability and the line $p=1/2$ where the dynamics becomes linear is also discussed  separately.

The time evolution equation for $f_+$ in $q\to \infty$ limit (Eq. (\ref{simplified mean field rate equation for q = infinity})) is given by
\begin{equation}
\label{rate equation for q tends to infinity}
 \frac{df_+}{dt} = h(f_+) = \frac{N(f_+)}{D(f_+)}.   
\end{equation}
where $N(f_+)=(1-2p)f_+^2+(2p-\alpha-1)f_++\alpha(1-p)$ and $D(f_+)=(1-p)+(2p-1)f_+$.

A regular fixed point $f_+^*$ satisfies $N(f_+^*)=0$ and $D(f_+^*)\neq 0$. Linearizing 
Eq. (\ref{rate equation for q tends to infinity}) by writing $f_+(t) = f_+^*+\delta(t)$ with $|\delta|\ll 1$ and using $N(f_+^*)=0$, we obtain to leading order,
\begin{equation}
 \frac{d\delta(t)}{dt} = h^{\prime}(f_+^*) \delta(t) \quad \text{with} \quad h^{\prime}(f_+^*) = \frac{N^{\prime}(f_+^*)}{D(f_+^*)}, 
\end{equation}
where $N^{\prime}(f_+^*)=2(1-2p)f_+^*+(2p-\alpha -1)$. Hence the infinitesimal deviation $\delta$ evolves exponentially as $\delta(t) = \delta(0)\exp(-|h^{\prime}(f_+^*)|t)$ with the local relaxation timescale associated with fixed point $f_+^*$ given by
\begin{equation}
 \tau_L(f_+^*) = \frac{1}{|h^{\prime}(f_+^*)|} = \frac{|D(f_+^*)|}{|N^{\prime}(f_+^*)|}.
\end{equation}
For $p\neq 1/2$, $N(f_+^*)$ is quadratic in $f_+^*$, so $N(f_+^*)=0$ gives two fixed points 
\begin{equation}
 f_+^*= r_{\pm} = \frac{-(2p-\alpha-1)\pm\sqrt{\Delta^{\prime}}}{2(1-2p)}, 
\end{equation}
with discriminant $ \Delta^{\prime} = (1-\alpha)^2 + 4p(1-p)(2\alpha-1) $. One can immediately see that for $0\leq p,\alpha\leq1$ we have $\Delta^{\prime}>0$. Therefore $\sqrt{\Delta^{\prime}}$ is real and we adopt the nonnegative branch $\sqrt{\Delta^{\prime}}\geq0$. In fact $\Delta^{\prime}=0$ occurs only for three special cases in the parameter combinations $(\alpha,p) = (0,1/2)$ or $(1,0)$ or $(1,1)$.

For $\sqrt{\Delta^{\prime}}>0$, the lower root $r_-$ is always the stable fixed point for which $h^{\prime}(r_-)=\frac{-2\sqrt{\Delta^{\prime}}}{1-\alpha+\sqrt{\Delta^{\prime}}}<0$. Hence the local relaxation timescale at $r_-$ is
\begin{equation}
\label{local timescale for q tends to infinity}
  \tau_L(r_-) = \frac{1-\alpha+\sqrt{\Delta^{\prime}}} {2\sqrt{\Delta^{\prime}}}.
\end{equation}
This compact expression for local timescale $\tau_L$ is valid for all $(\alpha ,p)$ except for the three special cases above where $\Delta^{\prime}=0$ and also not valid at the line $p=1/2$ for all non zero $\alpha$ where the dynamical equation for $f_+$ is linear in $f_+$ , so $r_{\pm}$ are no longer the roots of that equation and both the fixed points coalesce at $f_+^*=1/2$.

For $p=1/2$, $N(f_+)=\alpha(1/2-f_+),D(f_+)=1/2$ and $N^{\prime}(f_+)=-\alpha$. So the local relaxation timescale at the unique root $f_+^*=1/2$ is $\tau_{L}\big|_{\substack{p = 0.5 \\ \alpha \neq 0}}=\frac{1}{2\alpha}$ valid for all non zero $\alpha$.

If $\alpha=1$, then for any $0<p<1$, $\Delta^{\prime}=4p(1-p)>0$. Therefore from Eq. (\ref{local timescale for q tends to infinity}) one finds in this case $\tau_L(r_-)=1/2$.

Now consider the limiting cases at $p=0,1$ for both $\alpha \neq 1$ and  $\alpha=1$. For $\alpha\neq1$, the boundary points are not genuine fixed points. Indeed, at $p=0$ one has $N(f_+)=(f_+-1)(f_+-\alpha)$, $D(f_+)=1-f_+$ and therefore in this case fixed points can be either $\alpha$  or $1$. $f_+^*=\alpha$ is a valid stable fixed point as $h^{\prime}(\alpha)=\frac{N^{\prime}(\alpha)}{D(\alpha)}=-1<0$ whereas at $f_+^*=1$, $h(1)=\frac{N(1)}{D(1)}$ is naively $0/0$. To investigate further whether $f_+^*=1$ is a fixed point or not, we take the limit of $h(f_+^*)$ as $f_+^*\to1$. Taking the limit gives
\begin{equation}
\lim_{f_+^*\to1}h(f_+^*)=\lim_{f_+^*\to1} \frac{N(f_+^*)}{D(f_+^*)} = \alpha-1 \neq 0.
\end{equation}
So $f_+^*=1$ is not a fixed point for $p=0$. By symmetry $p\to 1-p$ and $f_+\to 1-f_+$, the same conclusion holds for $p=1$ that $\lim_{f_+^*\to 0} h(f_+^*)=1-\alpha\neq 0$, hence $f_+^*=0$ is also not a fixed point for $p=1$. Only when $\alpha=1$, the corner limit vanishes and therefore $f_+^*=1$ at $p=0$ and $f_+^*=0$ at $p=1$ become genuine fixed points.

When $\alpha=1$ and $p=0$, the only fixed point is the degenerate double root $f_+^*=1$ with $N(f_+)=(f_+-1)^2$ and $D(f_+)=1-f_+$. Hence $h^{\prime}(1)=N^{\prime}(1)/D(1)$ is naively $0/0$. To decide stability we take the limit of 
$h^{\prime}(f_+^*)$ as $f_+^*\to 1$. Taking the limit $f_+^*\to 1$ gives
\begin{equation}
  \lim_{f_+^*\to 1} h^{\prime}(f_+^*)=-2<0.
\end{equation}
So the degenerate fixed point $f_+^*=1$ for $p=0$ is stable as the linearization limit is negative. The local timescale in this case follows immediately
\begin{equation}
    \tau_L(1) = \lim_{f_+^*\to 1} \frac{1}{|h^{\prime}(f_+^*)|} = \frac{1}{2}.
\end{equation}
By symmetry, the case $\alpha=1,p=1$ maps to the previous one under the transformation 
$p\to 1-p,f_+\to 1-f_+$. In this case, the fixed point shifts to $f_+^*=0$, which again is a degenerate double root. Applying the same limiting analysis leads to the identical conclusion that the degenerate root is stable and the local relaxation timescale is $\tau_L(0)=1/2$.

\section{Emergence of $q$ independent global opinion dynamics of $f_+$ in large $q$ limit}

In this Appendix, we demonstrate that the dynamical evolution of the fraction of agents with positive opinion for the whole system ($f_+$) at time $t$, becomes independent of the group size $q$ in the limit $q\to \infty$.

The time evolution of $f_+(t)$ is governed by the master equation:
\begin{equation}
\label{General master equation}
 \frac{df_+(t)}{dt} = (1-2\alpha)\left[\sum_{n=0}^{q}p_{q+} \binom{q}{n} f_+^{n}(1-f_+)^{q-n}\right] + \alpha - f_+,
\end{equation}
where $p_{q+}(n) = \frac{np}{np + (q-n)(1-p)}$ is the probability that an agent adopts the positive opinion, given $n$ out of $q$ influencers are postive.

The sum which is inside the bracket of Eq. (\ref{General master equation}) is precisely the average value of $p_{q+}(n)$ which is $\langle p_{q+}(n)\rangle$.
 $Q(n) =\binom{q}{n} f_+^{n}(1-f_+)^{q-n} $ is the probability that, in a group of $q$ randomly selected agents, exactly $n$ of them have a positive opinion. This is precisely the binomial distribution of random variable $n$ with the mean $\mu = qf_+$ and variance $\sigma^2 =qf_+(1-f_+)$. So, the relative fluctuations $\frac{\sigma}{\mu} = \frac{\sqrt{qf_+(1-f_+)}}{qf_+} \sim \mathcal{O}\left(\frac{1}{\sqrt{q}}\right)\to 0$  as $q\to \infty$. This means  as $q\to \infty $, $Q(n)$ becomes sharply peaked around $n=qf_+$. Therefore a Taylor expansion of $p_{q+}(n)$ around $\mu$ up to $\mathcal{O}[(n-\mu)^2]$ yields
\begin{equation}
\label{taylor expansion of p_q+}
\langle p_{q+}(n)\rangle = p_{q+}(\mu) + \langle(n-\mu)\rangle p_q^{+\,\prime}(\mu) + \frac{1}{2} p_q^{+\,\prime\prime}(\mu) \langle (n-\mu)^2\rangle.   
\end{equation}
Now, putting $\langle(n-\mu)\rangle = 0$ and $\langle (n-\mu)^2\rangle = \sigma^2 = qf_+(1-f_+)$ and $p_q^{+\,\prime\prime}(\mu)\sim\mathcal{O}\left(1/q^2\right)$ in Eq. (\ref{taylor expansion of p_q+}), the correction term of $\mathcal{O}[(n-\mu)^2]$ is given by
\begin{equation}
 \frac{1}{2} p_q^{+\,\prime\prime}(\mu) qf_+(1-f_+) = \mathcal{O}\left (\frac{1}{q}\right).  
\end{equation}
Therefore in the large $q$ limit
\begin{equation}
 \langle p_{q+}(n)\rangle \approx \frac{f_+p}{f_+p+(1-f_+)(1-p)} + \mathcal{O}\left(\frac{1}{q}\right). 
\end{equation}
Substituting this in the master Eq. (\ref{General master equation}) yields
\begin{equation}
\label{q independence of dynmaical equation}
 \frac{df_+(t)}{dt} \approx (1-2\alpha)\left[\frac{f_+p}{f_+p+(1-f_+)(1-p)}+ \mathcal{O}\left(\frac{1}{q}\right)\right] + \alpha - f_+.    
\end{equation}
Hence in the limit $q\to \infty$, the dynamics of $f_+(t)$ becomes independent of $q$ with $\mathcal{O}[(n-\mu)^2]$ vanishing as $1/q$.

\bibliography{refs}

\end{document}